\documentclass[12pt,a4paper]{article}
\pdfoutput=1
\usepackage{jheppub}

\usepackage{amsmath,amssymb}
\usepackage[utf8]{inputenc}
\usepackage{graphicx}

\newcommand{\pv}{p_{t,\rm veto}}
\newcommand{\BSM}{\mathrm{BSM}}
\newcommand{\SM}{\mathrm{SM}}
\newcommand{\as}{\alpha_s}

\newcommand{\ptjv}{p_{\rm t, veto}}
\newcommand{\ETmiss}{{E\!\!\!/}_{\rm T}}
\newcommand{\ETmissrel}{{E\!\!\!/}_{\rm T,Rel}}
\newcommand{\proc}{{\rm i.s.}}
\newcommand{\nkll}{{\rm N}^{k}{\rm LL}}

\newcommand*{\jetvheto}{\texttt{JetVHeto}}
\newcommand*{\amcatnlo}{\texttt{aMC@NLO}}
\newcommand*{\mcfm}{\texttt{MCFM}}
\newcommand*{\mcfmre}{\texttt{MCFM-RE}}
\newcommand*{\mcherwig}{\amcatnlo{\tt +}\herwig}
\newcommand*{\mcpythia}{\amcatnlo{\tt +}\pythia}
\newcommand*{\aznlo}{\texttt{AZNLO}}
\newcommand*{\ctsix}{\texttt{CTEQ6L1}}
\newcommand*{\ctten}{\texttt{CT10}}

\newcommand*{\herwig}{\texttt{HERWIG}}
\newcommand*{\herwigseven}{\herwig\ \texttt{v7.1.0}}
\newcommand*{\pdfforlhc}{\texttt{PDF4LHC15}}
\newcommand*{\powheg}{\texttt{POWHEG}}
\newcommand*{\powherwig}{\powheg{\tt +}\herwig}
\newcommand*{\powpythia}{\powheg{\tt +}\pythia}
\newcommand*{\pythia}{\texttt{PYTHIA}}
\newcommand*{\pythiaeight}{\pythia\ \texttt{v8.230}}

\usepackage{color}
\definecolor{comment}{rgb}{0,0.3,0}
\definecolor{identifier}{rgb}{0.0,0,0.3}

\usepackage{listings}
\title{BSM $\mathbf{WW}$ production with a jet veto}
\date{\today}
\author[a]{Luke Arpino,}
\author[a]{Andrea Banfi,}
\author[a]{Sebastian J{\"a}ger,}
\author[b]{Nikolas Kauer}

\affiliation[a]{Department of Physics and Astronomy, University of Sussex, Brighton BN1 9QH, U.K.}
\affiliation[b]{Department of Physics, Royal Holloway, University of London, Egham Hill, Egham TW20 0EX, U.K.}

\emailAdd{l.arpino@sussex.ac.uk}
\emailAdd{a.banfi@sussex.ac.uk}
\emailAdd{s.jaeger@sussex.ac.uk}
\emailAdd{n.kauer@rhul.ac.uk}

\abstract{ We consider the impact on $WW$ production of the unique
  dimension-six operator coupling gluons to the Higgs field. In order
  to study this process, we have to appropriately model the effect of
  a veto on additional jets. This requires the resummation of large
  logarithms of the ratio of the maximum jet transverse momentum and
  the invariant mass of the $W$ boson pair. We have performed such
  resummation at the appropriate accuracy for the Standard Model (SM)
  background and for a signal beyond the SM (BSM), and devised a
  simple method to interface jet-veto resummations with fixed-order
  event generators. This resulted in the fast numerical code \mcfmre,
  the Resummation Edition of the fixed-order code \mcfm. We compared
  our resummed predictions with parton-shower event generators and
  assessed the size of effects, such as limited detector acceptances,
  hadronisation and the underlying event, that were not included in
  our resummation. We have then used the code to compare the
  sensitivity of $WW$ and $ZZ$ production at the HL-LHC to the
  considered higher-dimension operator. We have found that $WW$ can
  provide complementary sensitivity with respect to $ZZ$, provided one
  is able to control theory uncertainties at the percent-level. Our
  method is general and can be applied to the production of any colour
  singlet, both within and beyond the SM.}

\begin{document}
\maketitle

\section{Introduction}
\label{sec:intro}

Di-boson production at the Large Hadron Collider constitutes a
promising window into physics beyond the SM. This is particularly true
for di-boson pairs with high invariant mass, which have been already
probed by a number of recent experimental
analyses~\cite{Aaboud:2018jqu,Aaboud:2017gsl,Aaboud:2017qkn,Aaboud:2018puo,Aaboud:2017itg,Aaboud:2017eta,Aaboud:2017fgj,Aaboud:2019lxo,Sirunyan:2017zjc,Sirunyan:2018egh,Sirunyan:2019twz}. On
the one hand, their production through gluon fusion receives
contributions from an off-shell Higgs
boson~\cite{Glover:1988fe,Kauer:2012hd,Kauer:2013qba}. In particular,
the interference of the contribution of an off-shell Higgs boson and
di-boson continuum background makes it possible to access the Higgs
width in a model-independent way~\cite{Caola:2013yja}. On the other
hand, contact interactions arising from higher-dimensional effective
field theory
operators~\cite{Buchmuller:1985jz,Grzadkowski:2010es,Giudice:2007fh,Elias-Miro:2013mua,Harlander:2013oja}
could give rise to spectacular effects in the tails of di-boson
differential distributions, due to the fact that their contribution
increases with energy. Technically, in the SM, di-boson production via
gluon fusion is a loop-induced process. At low di-boson invariant
masses, top quarks in the loops behave as very heavy particles, thus
giving rise to effective contact interactions. At high invariant
masses, the two bosons probe virtualities that are much larger than
the masses of the top quarks running in the loops, hence suppressing
their contribution and enhancing the effect of BSM contact
interactions. Such a feature has been already used to constrain the
coefficient of a number of higher-dimensional operators, see
e.g.~\cite{Ellis:2018gqa} for a recent study.

In this article we restrict ourselves to considering the unique
dimension-six operator coupling gluons to the Higgs boson, given
by~\cite{Grzadkowski:2010es}
\begin{equation}
  \label{eq:BSM-operator}
   \mathcal{L} \supset \frac{c_{gg}}{\Lambda^2} G^a_{\mu\nu} G^{a,\mu\nu} \phi^\dagger \phi ,
\end{equation}
with $G^a_{\mu\nu}$ the gluon field strength and $\phi$ the Higgs
field. This operator can be used to represent contributions to SM
Higgs production from particles with mass of order $\Lambda\gg m_H$.
This operator has previously been considered in high-invariant-mass
$ZZ$ production with a fully leptonic final state
in~\cite{Azatov:2014jga,Buschmann:2014sia}. However, the leptonic
final state for $WW$ has larger cross section and so $WW$ could give
complementary or better sensitivity than leptonic final states for
$ZZ$. However, in $WW$ production, a tight jet veto is employed by
experiments to suppress background from top-pair production. Such a
veto ``forbids'' the radiation of jets from the initial-state partons,
with the effect of suppressing not only the background, but also the
operator-mediated signal. In the present case, the signal occurs
through gluon fusion, whereas $WW$ production is mainly driven by
quark-antiquark annihilation. Since gluons radiate more than quarks,
one expects the suppression due to a jet veto to be stronger for the
signal than for the background. It is therefore important to address
the general question of how BSM searches with $WW$ production compare
to $ZZ$ in the presence of a jet veto.\footnote{In fact, a supposed
  discrepancy of the total $WW$ cross section from SM
  predictions~\cite{ATLAS:2012mec,Chatrchyan:2013yaa,Chatrchyan:2013oev}
  could be partly ascribed to mismodelling of jet-veto
  effects~\cite{Jaiswal:2014yba,Monni:2014zra,Becher:2014aya,Re:2018vac,Jaiswal:2015vda}.}

The aim of this paper is to quantify in a simple way how the
significance of such a BSM signal is affected by the presence of a jet
veto. The same procedure can be applied to any BSM scenario that
modifies the production rate of a colour singlet, for instance
dimension-8 operators~\cite{Bellm:2016cks}. A similar
study~\cite{Moult:2014pja} investigates the impact of a jet veto in
the determination of the Higgs width using interference.  To be more
specific, we veto all jets that have a transverse momentum (with
respect to the beam axis) above $\ptjv$. First we observe that, at the
level of the matrix element squared, a generic BSM signal mediated by
a single higher-dimensional operator consists of an interference piece
and a quadratic piece:
\begin{equation}
   |M_{\rm SM}|^2 + 2 {\rm Re} (M_{\rm SM}^* M_{\rm BSM}) + |M_{\rm BSM}|^2.
\end{equation}
The last piece is of higher order $1/\Lambda^4$. Therefore, if the
interference piece is not suppressed or vanishing for some reason,
then, to a first approximation, we can neglect it relative to the
$1/\Lambda^2$ interference piece.\footnote{Note that, if we consider
  more than one higher-dimensional operator, there are possibly other
  BSM effects of order $1/\Lambda^3$ or $1/\Lambda^4$ in the
  interference piece in general, which might still compete with or
  dominate over the quadratic piece.} The presence of a jet veto
induces large logarithms of the ratio of $\ptjv$ and the invariant
mass of the $WW$ pair $M_{WW}$. Such logarithms arise at all orders in
QCD, and originate from vetoing soft-collinear parton
emissions. Considering just the leading logarithms, and neglecting the
quadratic piece $|M_{\rm BSM}|^2$, the deviation of a BSM signal that
proceeds from gluon fusion from the SM prediction is approximately
given by
\begin{equation}
\label{eq:dMsq-BSM}
   \mathcal{L}_{gg} (M_{WW})\times 2 {\rm Re} (M_{\rm SM}^* M_{\rm BSM})\, e^{-2 C_A \frac{\alpha_s}{\pi}\ln^2\left(\frac{M_{WW}}{\ptjv}\right)}\,,
\end{equation}
where $C_A=3$, $\alpha_s$ is the strong coupling, and
$\mathcal{L}_{gg} (M_{WW})$ is the gluon-gluon luminosity
corresponding to a partonic centre-of-mass energy equal to
$M_{WW}$. The effect of the jet veto is an exponential (Sudakov)
suppression with respect to a naive Born-level estimate. Note also
that, for fixed $\ptjv$, such a suppression becomes more and more
important, the higher the invariant mass of the $WW$ pair. This is
precisely where the contribution of the BSM operator in
eq.~\eqref{eq:BSM-operator} has the most impact on the signal. For the
SM background, dominated by quark-antiquark annihilation, we have
instead a contribution proportional to
\begin{equation}
\label{eq:Msq-SM}
   \mathcal{L}_{q\bar q} (M_{WW})\times |M_{\rm SM}|^2 \, e^{-2 C_F \frac{\alpha_s}{\pi}\ln^2\left(\frac{M_{WW}}{\ptjv}\right)}\,,
\end{equation}
with $C_F=4/3$ and $\mathcal{L}_{q\bar q} (M_{WW})$ the
quark-antiquark luminosity. The relative deviation from the SM can be
obtained by integrating eqs.~\eqref{eq:dMsq-BSM} and~\eqref{eq:Msq-SM}
over the appropriate phase space. Note that, for a fixed value of
$M_{WW}$, the exponential encoding jet-veto effects factorise
completely. Therefore, the relative deviation from the SM in the
presence of the jet-veto is different from that obtained with a
Born-level calculation by a factor
\begin{equation}
  \label{eq:Sudakov-ratio}
  e^{-2 (C_A-C_F) \frac{\alpha_s}{\pi}\ln^2\left(\frac{M_{WW}}{\ptjv}\right)}\,.
\end{equation}
For $\alpha_s=0.1, M_{WW}= 1\,\mathrm{TeV}, \ptjv=20\,\mathrm{GeV}$,
the above factor is about 0.2. Therefore, despite the gain in the
number of events one has in $WW$ production with respect to $ZZ$, the
significance of the signal might be reduced due to jet-veto
effects. This is why it is crucial to have an estimate of jet-veto
effects that is as accurate as possible.

The first question we address is what accuracy we can aim for in the
description of a BSM signal and a QCD background involving the
production of a colour singlet. In the absence of large jet-veto
corrections, a generic BSM signal can be predicted at Born-level, or
leading order (LO), in QCD, whereas any QCD background is nowadays
known at least at next-to-leading order (NLO). In the presence of a
jet veto, the production of a system of invariant mass $M$ is affected
by logarithms of the ratio $\ptjv/M$, which make fixed-order
predictions unreliable. After the all-order resummation of
$\ln(\ptjv/M)$, the differential cross-section $d\sigma (\ptjv)/dM^2$
with no jets with a transverse momentum above $\pv$ can be written in
the form\footnote{The general expression in eq.~\eqref{eq:sigma-res}
  holds because the transverse momentum of the leading jet has the
  property of recursive infrared and collinear (rIRC)
  safety~\cite{Banfi:2004yd}.}
\begin{equation}
  \label{eq:sigma-res}
  \frac{d\sigma(\ptjv)}{dM^2} = \frac{d\sigma_0}{dM^2}\,e^{Lg_1(\alpha_s L)}\left(G_2(\alpha_s L)+\alpha_s G_3(\alpha_s L)+\dots\right)\,,\quad L\equiv \ln\frac{M}{\pv}\,.
\end{equation}
with $d\sigma_0/dM^2$ the corresponding LO cross section. The above
expression is meaningful for $\alpha_s \ln(M/\pv)\sim 1$ and misses
terms that vanish as powers of $\pv/M$ (possibly enhanced by
logarithms). The leading logarithmic (LL) contributions exponentiate
giving rise to the function $g_1(\alpha_s L)$, with $\alpha_s$
evaluated at a renormalisation scale of order $M$. Next-to-leading
logarithmic terms (NLL) factorise from LL ones, and are embedded in
the function $G_2(\alpha_s L)$.  Next-to-next-to-leading logarithmic
(NNLL) contributions, resummed by $G_3(\alpha_s L)$, are of relative
order $\alpha_s$ with respect to NLL ones, and similarly one can
define higher logarithmic accuracy. The knowledge of NLO correction to
a QCD background process gives access to all ingredients to compute
$G_3$, i.e.\ to achieve NNLL accuracy, whereas the lack of knowledge
of corrections of relative order $\alpha_s$ to a generic BSM process
implies that the best accuracy one can aim at for such processes is
NLL. Therefore, from the point of view of the accuracy of the
resummation, having LO makes it possible to reach NLL accuracy,
whereas NLO gives access to NNLL accuracy.

The most widely used method to estimate jet-veto effects are Monte
Carlo event generators, which simulate the contribution of multiple
soft-collinear QCD emissions. Although very flexible, these tools
cannot formally guarantee more than LL accuracy, and at the moment
require a considerable amount of tuning to reliably describe
observables, like the cross section with a jet-veto in
eq.~\eqref{eq:sigma-res}, sensitive to QCD radiation from the initial
state (see e.g.~\cite{Aad:2015auj} for a recent study). In order to
have more accurate predictions, one needs to consider analytical
resummations.

Jet-veto effects in the production of a colour singlet have been
computed at NNLL accuracy in QCD~\cite{Banfi:2012jm} and in
soft-collinear effective theory
(SCET)~\cite{Becher:2013xia,Stewart:2013faa}.\footnote{The resummed
  predictions of refs.~\cite{Becher:2013xia,Stewart:2013faa} include
  all constants multiplying the resummation at order $\alpha_s^2$,
  which are formally N$^3$LL in the counting of
  eq.~(\ref{eq:sigma-res}). However, the resummations of
  refs.~\cite{Becher:2013xia,Stewart:2013faa} do not account for all
  N$^3$LL effects, and their accuracy is labelled NNLL$'$. The same
  accuracy can be obtained by matching a NNLL resummation with exact
  NNLO. Although feasible, this is beyond the scope of the present
  work.} The calculation of~\cite{Banfi:2012jm} is implemented for
Higgs and $Z$-boson production, inclusive in all decay products, in
the program \jetvheto~\cite{jetvheto}. A NNLL resummation in SCET
using the results of ref.~\cite{Becher:2013xia} has been implemented
in \amcatnlo\ for the production of a generic colour singlet, fully
exclusive in its decay products~\cite{Becher:2014aya}. This
implementation has been used to estimate jet-veto effects in $WW$
production~\cite{Becher:2014aya} in the SM, and for hypothetical $Z'$
and $W'$ bosons~\cite{Fuks:2017vtl}. The specificity of this
implementation is the way it handles the so-called ``beam
functions''. These contain convolutions of appropriate coefficient
functions with parton distibutions, and are a general feature of NNLL
resummations with hadrons in the initial state. In
ref.~\cite{Becher:2014aya}, beam functions are precomputed and
tabulated so as to replace traditional parton distribution
functions. In this article, we discuss an alternative approach that
implements the QCD resummation of ref.~\cite{Banfi:2012jm}, fully
exclusive in the decay products of the colour singlet, in a way that
is not tied to a specific event generator (e.g.\ \amcatnlo), but that
requires minimal and simple modifications of the setup that is already
available in any NLO QCD program.
The starting point is to observe that, in eq.~\eqref{eq:sigma-res},
the factor multiplying leading logarithms is in fact a new
perturbative series, whose coefficients are functions of $\alpha_s L$.
As stated previously, NLL corrections have the same structure as
Born-level contributions, while NNLL corrections closely resemble NLO
contributions. Therefore, NLL resummation could just be obtained by an
event-by-event reweighting of a Born-level generator by keeping only
the functions $g_1$ and $G_2$ in eq.~\eqref{eq:sigma-res}. This is
enough to estimate jet-veto effects to the BSM production of a colour
singlet. Including NNLL corrections, needed for a precise estimate of
the corresponding SM background, is also possible in a general way.
In fact, resummation effects originate from soft and/or collinear
emissions in such a way that NNLL corrections share the same phase
space with Born-level contributions, but are of relative order
$\alpha_s$. In all NLO calculations there is always a contribution
that lives in the same phase space as the Born, and is of relative
order $\alpha_s$. This is the subtraction term that cancels the
infrared singularities of virtual corrections. Therefore, to implement
NNLL effects, we can just modify the appropriate subtraction term in
the NLO event generator. Having done this, all other NNLL effects
factorise, and can be accounted for by an event-by-event reweighting,
so as to reproduce eq.~\eqref{eq:sigma-res}. The whole procedure
requires generating Born-level events only, and hence is much faster
than a full NLO calculation. As will be clear later, the same approach
can be used to interface resummations in SCET, provided one is able to
rewrite results in terms the functions $G_2$ and $G_3$ in
eq.~\eqref{eq:sigma-res}.

In the following two sections we give a detailed description of this
procedure for the specific case of BSM effects induced by the operator
in eq.~(\ref{eq:BSM-operator}).
In section~\ref{sec:signal}, we study the effect of such an operator
on $WW$ production with a jet veto. As discussed above, this operator
induces a modification of the cross section of $WW$ production through
gluon fusion. We denote the (differential) cross section for gluon
fusion, potentially including an additional BSM contribution, with
$d\sigma_{gg}$. The main result of this section is a recipe to compute
cross sections for $WW$ production with a jet veto at NLL accuracy,
fully exclusive in the decay products of the $W$ bosons.
In section~\ref{sec:background} we compute the cross section for the
dominant contribution to the SM background, which is $WW$ production
via quark-antiquark annihilation, again in the presence of a jet
veto. We denote the cross-section for this process with
$d\sigma_{q\bar q}$, and compute exclusive cross sections in the decay
products of the $W$ bosons, while resumming $\ln(M_{WW}/\ptjv)$ at
NNLL accuracy. The main result of this section is a general recipe to modify
a NLO event generator for the production of any colour singlet so that
it produces resummed cross-section with a jet veto at NNLL accuracy.
In section~\ref{sec:results} we present some numerical results for a
simplified model derived from the Lagrangian in
eq.~(\ref{eq:BSM-operator}), corresponding to a realistic experimental
setup. We compare our resummed predictions with parton-shower event
generators, and assess the size of effects, such as limited detector
acceptances, hadronisation and the underlying event, that are not
included in our resummation.
In section~\ref{sec:sensitivity} we perform some basic sensitivity
studies to investigate the exclusion potential of the HL-LHC for the
parameters of the simplified model of section~\ref{sec:results}.
Finally, section~\ref{sec:conclusions} presents our conclusions. 

\section{Gluon fusion (including BSM effects)}
\label{sec:signal}

Let us first consider $WW$ production via gluon fusion, possibly with
a modification of the amplitude induced by the BSM operator in
eq.~(\ref{eq:BSM-operator}). For simplicity, we consider here the
decays $WW\to e^+ \nu_e \mu^- \bar \nu_\mu$ and
$WW\to e^- \bar \nu_e \mu^+ \nu_\mu$. As explained in the
introduction, if we impose that all jets have a transverse momentum
below a threshold value $\ptjv$, the distribution in $M_{WW}^2$,
differential in the phase space of the leptons, is affected by the
presence of large logarithms $\ln(M_{WW}/\ptjv)$, that have to be
resummed to all orders to obtain sensible theoretical
predictions. Specifically, we consider jets obtained by applying the
anti-$k_t$ algorithm~\cite{Cacciari:2008gp} with a given radius
$R$. At NLL accuracy, the best we can achieve for gluon fusion, the
aforementioned observable is given by~\cite{Banfi:2012yh,Banfi:2012jm}
\begin{equation}
  \label{eq:dsigma-BSM}
  \frac{d\sigma^{\rm NLL}_{gg}(\pv)}{d\Phi_{\rm leptons}dM^2_{WW}} = {\mathcal
      L}_{gg}^{(0)}(L,M_{WW}) \,e^{ Lg_1(\as  L)+g_2(\as  L)}\,,
\end{equation}
where $L=\ln(M_{WW}/\ptjv)$, $\alpha_s=\alpha_s(M_{WW})$, and explicit
expressions for the functions $g_1(\as L)$ and $g_2(\as L)$ can be
found, for instance, in ref.~\cite{Banfi:2012jm}. In particular, they
are the same for any colour singlet that is produced via gluon fusion
(e.g.\ Higgs production). Note that, at NLL accuracy, the resummed
distribution in eq.~(\ref{eq:dsigma-BSM}) does not depend on the
radius $R$ of the jets~\cite{Banfi:2012yh}.

The phase space of the leptons is given by
\begin{equation}
  \label{eq:phase-space}
  d\Phi_{\rm leptons}=\frac{d^3 \vec p_e}{(2\pi)^3 2 E_e}\frac{d^3 \vec p_{\nu_e}}{(2\pi)^3 2 E_{\nu_e}}\frac{d^3 \vec p_\mu}{(2\pi)^3 2 E_
    \mu} \frac{d^3 \vec p_{\nu_\mu}}{(2\pi)^3 2 E_{\nu_\mu}}(2\pi)^4\delta\left(p_e+p_\mu+p_{\nu_e}+p_{\nu_\mu}-p_1-p_2\right)\,,
\end{equation}
with $p_{\ell}=(E_\ell,\vec p_\ell)$ is the four-momentum of lepton
$\ell=e,\mu,\nu_e,\nu_\mu$, and $p_i=x_i P_i, i=1,2$ are the momenta
of the incoming partons, carrying each a fraction $x_i$ of the
incoming proton momentum $P_i$.

Last, we have a process dependent ``luminosity'' factor
${\mathcal L}_{gg}^{(0)}$, given by\footnote{Note that, since
  jet-veto measurements do not keep track of any correlation between
  the angle of the jet and the outgoing leptons, the logarithmically
  enhanced contributions due to the helicity of incoming gluons
  described in~\cite{Catani:2010pd} are not present in our case, so
  eq.~\eqref{eq:LBSM} is valid as is. }

\begin{multline}
 \label{eq:LBSM}
 {\mathcal L}_{gg}^{(0)}(L,M_{WW}) = \int dx_1 dx_2 \, |M_{\SM}^{(gg)}+ M^{(gg)}_{\rm BSM}|^2 \,\delta(x_1 x_2 s - M_{WW}^2)\times \\ \times
  f_g\!\left(x_1, \ptjv\right)f_g\!\left(x_2, \ptjv\right)\,.
\end{multline}
The two main ingredients entering ${\mathcal L}_{gg}^{(0)}$ are:
\begin{itemize}
\item the SM amplitude $M^{(gg)}_{\SM}$ for the production of a $WW$
  pair (and its decay products) through gluon fusion, which can be
  supplemented with an additional contribution $M^{(gg)}_{\rm BSM}$
  accounting for BSM effects;
\item the gluon density in the proton $f_g(x,\mu_F)$ at the
  factorisation scale $\mu_F=\ptjv$. This value of $\mu_F$ reflects
  the fact that the factorisation scale is the highest scale up to
  which the considered observable is inclusive with respect to
  multiple collinear emissions from the initial-state partons. Since
  all collinear emissions with a transverse momentum above $\ptjv$ are
  vetoed, the factorisation scale has to be $\ptjv$ (see
  e.g.~\cite{Banfi:2004yd} for a formal derivation).
\end{itemize}
By comparing eq.~\eqref{eq:dsigma-BSM} to eq.~(\ref{eq:sigma-res}),
we obtain the function $G_2(\alpha_s L)$ resumming all NLL
contributions:
\begin{equation}
  \label{eq:G2-BSM}
  G_2(\alpha_s L) = \frac{{\mathcal L}_{gg}^{(0)}(L,M_{WW})}{{\mathcal L}_{gg}^{(0)}(0,M_{WW})}\, e^{g_2(\alpha_s L)}\,.
\end{equation}

So far, with the exception of ref.~\cite{Becher:2014aya}, such
resummations have been obtained by devising process-dependent codes
that produce numerical results for
${\mathcal L}_{gg}^{(0)}(L,M_{WW})$. For instance, the program
\jetvheto~\cite{jetvheto} returns NNLL resummations integrated over the
full phase space of the decay products of a Higgs or a $Z$ boson.
However, the luminosity in eq.~\eqref{eq:LBSM} can be obtained by
running any Born-level event generator. In fact, any such program will
compute a Born-level cross-section in $WW$ production via gluon fusion
(possibly with BSM contributions) starting from the formula:
\begin{multline}
  \label{eq:dsigma-LO}
  \frac{d\sigma^{(0)}_{gg}}{d\Phi_{\rm leptons} dM^2_{WW}}=\int dx_1
  dx_2 \, |M_{\SM}^{(gg)}+ M^{(gg)}_{\rm BSM}|^2\,\delta(x_1 x_2 s -
  M_{WW}^2)\times \\ \times f_g\!\left(x_1,
    \mu_F\right)f_g\!\left(x_2,\mu_F\right)={\mathcal
    L}_{gg}^{(0)}(0,M_{WW})\,,
\end{multline}
where $\mu_F$ here is the default factorisation scale in the
considered Born-level generator. Therefore, to obtain the differential
distribution in eq.~(\ref{eq:dsigma-BSM}), it is enough to set that
factorisation scale $\mu_F$ to $\ptjv$, and multiply the weight of
each phase-space point by $\exp[Lg_1(\alpha_s L)+g_2(\alpha_s L)]$.
Note that, if the programs returns event files with information on
$M_{WW}$ for each event, or if one produces histograms binned in
$M_{WW}$, the reweighting can be performed without any need to touch
the Born-level generator code.

\section{Quark-antiquark annihilation (SM only)}
\label{sec:background}

Since SM background processes are typically known at least to NLO, in
the presence of a jet veto, the SM cross-section for $WW$ production
can be computed at NNLL accuracy. The corresponding NNLL resummed
expression is given by
\begin{multline}
  \label{eq:fullxsc}
  \frac{d\sigma^{\rm NNLL}_{q\bar q}(\pv)}{d\Phi_{\rm leptons} dM_{WW}^2}= \left( {\mathcal
      L}_{q\bar q}^{(0)}(L,M_{WW})+{\mathcal L}_{q\bar q}^{(1)}(L,M_{WW})\right) \times
  \\ \times
  \left (1+\mathcal{F}_{\text{clust}}(R)+\mathcal{F}_{\text{correl}}(R)\right)
  \times e^{ Lg_1(\as  L)+g_2(\as  L)+\frac{\as}{\pi} g_3(\as L)}\,,
\end{multline}
where again $L=\ln(M_{WW}/\ptjv)$, $\alpha_s=\alpha_s(M_{WW})$, and
$d\Phi_{\rm leptons}$ is the lepton phase space defined in
eq.~(\ref{eq:phase-space}). The functions $g_1,g_2$ and $g_3$ are
reported in~\cite{Banfi:2012jm}, and are the same as for Drell-Yan
production. The dependence on the jet radius $R$ appears for the first
time at NNLL accuracy in the functions
$\mathcal{F}_{\text{clust}}(R)$, $\mathcal{F}_{\text{correl}}(R)$,
whose explicit expressions can be found in~\cite{Banfi:2012yh}.

At NNLL accuracy we have two process-dependent ``luminosities''
$\mathcal{L}_{q\bar q}^{(0)}$ and $\mathcal{L}_{q\bar q}^{(1)}$. The
luminosity $\mathcal{L}_{q\bar q}^{(0)}$ is the analogue of
$\mathcal{L}_{gg}^{(0)}$ of eq.~(\ref{eq:LBSM}), this time for a
$q\bar q$ initiated process:
\begin{align}
 \label{eq:L0-qq}
  & {\mathcal L}_{q\bar q}^{(0)}(L,M_{WW}) = \sum_{i,j}\int dx_1 dx_2 |M^{(q\bar q)}_{ij}|^2\delta(x_1 x_2 s - M_{WW}^2)
    f_i\!\left(x_1, \ptjv\right)f_j\!\left(x_2, \ptjv\right)\,.
\end{align}
The only difference with respect to $\mathcal{L}_{gg}^{(0)}$ is the
LO SM amplitude $M^{(q\bar q)}_{ij}$, which is different from zero only if
$i,j$ is a quark-antiquark pair with the same flavour. 

At NNLL accuracy we need to add the luminosity
$\mathcal{L}_{q\bar q}^{(1)}$, which is of relative order $\alpha_s$ with
respect to $\mathcal{L}_{q\bar q}^{(0)}$, and is given by
\begin{align}
\label{eq:L1-qq}
&   {\mathcal L}_{q\bar q}^{(1)}(L,M_{WW})  =  \sum_{i,j}\int
                           dx_1 dx_2 |M^{(q\bar q)}_{ij}|^2 \delta(x_1 x_2 s -
                M_{WW}^2) \times  \\ &
                                       \qquad\qquad\qquad\times
 \bigg[f_i\!\left(x_1, \ptjv\right)
  f_j\!\left(x_2, \ptjv\right)\frac{\alpha_s(M_{WW})}{2\pi}{\cal H}^{(1)} +\notag\\
  &\frac{\alpha_s(\ptjv)}{2\pi}\sum_{k}\bigg(
  \int_{x_1}^1\frac{dz}{z} C_{ik}^{(1)}(z)
  f_k\!\left(\frac{x_1}{z}, \ptjv\right)
  f_j\!\left(x_2, \ptjv\right) +\{(x_1,i)\,\leftrightarrow\,(x_2,j)\}\bigg)\, \bigg]\,. 
\end{align}
Here new ingredients appear:
\begin{itemize}
\item one-loop virtual corrections to $WW$ production. They are included in the term ${\cal H}^{(1)}$, the coefficient of $\alpha_s(M_{WW})$;
\item coefficient constants arising from real collinear
  radiation. They are included in the terms $C_{ik}^{(1)}(z)$, whose
  explicit expressions can be found in ref.~\cite{Banfi:2012jm}, and are
  the same as for Drell-Yan production. They multiply
  $\alpha_s(\ptjv)$, which reflects the fact that the characteristic
  scale of collinear radiation in jet-veto cross sections is $\ptjv$.
\end{itemize}
With reference to eq.~(\ref{eq:sigma-res}), the function $G_2$
resumming NLL contributions is
\begin{equation}
  \label{eq:G2-SM}
  G_2(\alpha_s L) = \frac{{\mathcal L}_{q\bar q}^{(0)}(L,M_{WW})}{{\mathcal L}_{q\bar q}^{(0)}(0,M_{WW})}\, e^{g_2(\alpha_s L)}\,,
\end{equation}
whereas the function $G_3$ resumming NNLL contributions is
\begin{multline}
  \label{eq:G3-SM}
  G_3(\alpha_s L) = \frac{e^{g_2(\alpha_s L)}}{\alpha_s{\mathcal L}_{q\bar q}^{(0)}(0,M_{WW})}
  \left[{\mathcal L}_{q\bar q}^{(1)}(L,M_{WW}) \right. \\ \left. +{\mathcal L}_{q\bar q}^{(0)}(L,M_{WW})\left(\mathcal{F}_{\text{clust}}(R)+\mathcal{F}_{\text{correl}}(R)+
      \frac{\alpha_s}{\pi} g_3(\alpha_s L)\right)\right]\,.
\end{multline}

As explained in the previous section, the function
${\mathcal L}_{q\bar q}^{(0)}$ can be obtained from an appropriate
Born-level program. The function ${\mathcal L}_{q\bar q}^{(1)}$ instead
represents a correction to ${\mathcal L}_{q\bar q}^{(0)}$ of relative
order $\alpha_s$, that cannot be obtained from a LO calculation. A
viable possibility to perform NNLL resummation would be to modify
eq.~(\ref{eq:LBSM}) so that it includes the convolutions over the
variable $z$ in eq.~\eqref{eq:L1-qq}, and implement the modification in a
Born-level generator. This is the approach taken in
ref.~\cite{Becher:2014aya}, and in some way underlying the current
implementation of the \jetvheto\ program~\cite{jetvheto}. Here we want
to present an alternative procedure. First, let us consider how the
NLO $WW$ cross section is calculated in a NLO event generator:
\begin{equation}
  \label{eq:NLOxsc}
  \frac{d\sigma^{\rm NLO}_{q\bar q}(\pv)}{d\Phi_{\rm leptons} dM_{WW}^2}= 
\frac{d\sigma_{q\bar q}^{(0)}}{d\Phi_{\rm leptons} dM_{WW}^2}+\frac{d\sigma_{q\bar q,v+ct}^{(1)}}{d\Phi_{\rm leptons} dM_{WW}^2}+\frac{d\sigma_{q\bar q,r}^{(1)}}{d\Phi_{\rm leptons} dM_{WW}^2}\,.
\end{equation}
The first term in the sum is the LO SM cross section
$d\sigma_{q\bar q}^{(0)}/d\Phi_{\rm leptons} dM_{WW}^2= {\mathcal
  L}_{q\bar q}^{(0)}(0,M_{WW})$.
The last term, $d\sigma_{q\bar q,r}^{(1)}/d\Phi_{\rm leptons}  dM_{WW}^2$, represents
NLO corrections coming from the emission of an extra parton. They
include the counterterms needed to ensure their finiteness in four
space-time dimensions. The second term,
$d\sigma_{q\bar q,v+ct}^{(1)}/d\Phi_{\rm leptons} dM_{WW}^2$, gives NLO corrections
arising from the sum of virtual corrections, and the counterterms
integrated over the full extra-parton phase space. This contribution lives in
the same phase-space as the Born contribution, and is of relative
order $\alpha_s$. It has the form
\begin{equation}
  \label{eq:v+ct}
  \begin{split}
    &\frac{d\sigma_{q\bar q,v+ct}^{(1)}}{d\Phi_{\rm leptons} dM_{WW}^2} =
    \frac{\alpha_{s}(\mu_R)}{2\pi}\sum_{i,j}\int dx_1 dx_2 |M^{(q\bar q)}_{ij}|^2
    \delta(x_1 x_2 s - M_{WW}^2) \bigg[f_i\!\left(x_1, \mu_F\right)
    f_j\!\left(x_2, \mu_F\right)\tilde {\cal H}^{(1)} +\\
    &\sum_{k}\bigg( \int_{x_1}^1\frac{dz}{z} \tilde C_{ik}^{(1)}(z)
    f_k\!\left(\frac{x_1}{z}, \mu_F\right) f_j\!\left(x_2,\mu_F\right)
    +\{(x_1,i)\,\leftrightarrow\,(x_2,j)\}\bigg)\, \bigg]\,.
  \end{split}
\end{equation}
In the above equation, $\mu_R,\mu_F$ are the renormalisation and
factorisation scales used by the NLO generator,
$\tilde {\cal H}^{(1)}$ represents virtual corrections to
$q\bar q\to WW$, and $\tilde C^{(1)}_{ik}(z)$ the integrated
counterterms. The explicit expressions of $\tilde {\cal H}^{(1)}$ and
$\tilde C^{(1)}_{ik}(z)$ depend on their actual implementation in the
NLO generator, in particular on the employed subtraction
scheme. However, the form of eq.~\eqref{eq:v+ct} is the same as that
of the NNLL luminosity ${\mathcal L}_{q\bar q}^{(1)}(L,M_{WW})$ in
eq.~(\ref{eq:L1-qq}). Therefore, by comparing
eqs.~(\ref{eq:L1-qq}) and~\eqref{eq:v+ct}, we find that
${\mathcal L}_{q\bar q}^{(1)}(L,M_{WW})$ can be implemented in a NLO event
generator by performing the replacements
\begin{equation}
  \label{eq:nlo-replacements}
  \begin{split}
    & \frac{\alpha_{s}(\mu_R)}{2\pi} \tilde {\cal H}^{(1)} \to \frac{\alpha_{s}(M_{WW})}{2\pi} \mathcal{H}^{(1)}\,,\\
    & \frac{\alpha_{s}(\mu_R)}{2\pi} \tilde C^{(1)}_{ik}(z) \to \frac{\alpha_{s}(\ptjv)}{2\pi} C^{(1)}_{ik}(z)\,,
  \end{split}
\end{equation}
and by evaluating the parton distribution functions at the
factorisation scale $\mu_F=\ptjv$. Finally, in order to obtain the
resummed distribution in eq.~(\ref{eq:fullxsc}), we need to reweight
each phase space point by
\begin{equation}
  \label{eq:rescale-NNLL}
\left
  (1+\mathcal{F}_{\text{clust}}(R)+\mathcal{F}_{\text{correl}}(R)\right)
e^{ Lg_1(\as L)+g_2(\as L)+\frac{\as}{\pi} g_3(\as L)}\,.
\end{equation}
This rescaling can also be performed when constructing histograms, as
long as one has access to $M_{WW}$ for each bin, or for each event in
an event record.

We have implemented this procedure in the code \mcfmre,\footnote{Available on request.}
a suitable modification of the NLO program \mcfm~\cite{mcfm}. The
actual implementation is richer than what has been discussed so far,
because it allows a user to change the default renormalisation and
factorisation scales, and contains additional features. Since these
details are not relevant for a general discussion, we have omitted
them here. The interested reader is referred to
appendix~\ref{sec:formulae} for the actual formulae we implement, and
to appendix~\ref{sec:mcfm-re} for a short manual of the code.

In the following two sections, we use this implementation to produce
numerical results and sensitivity studies for an explicit BSM model.

\section{Numerical results}
\label{sec:results}

Let us discuss first our results for $WW$ production via $q\bar q$
annihilation. We consider $W$ pairs produced at the LHC with
$\sqrt s = 13\,\mathrm{TeV}$, specifically
$W^+ W^-\to e^+ \nu_e \mu^- \bar \nu_\mu$ and
$W^+ W^-\to \mu^+ \nu_\mu e^- \bar \nu_e$, and select the final state
according to a simplified version of the experimental cuts of
ref.~\cite{Aaboud:2017qkn}, reported in table~\ref{tab:cuts}. Jets are
reconstructed according to the anti-$k_t$
algorithm~\cite{Cacciari:2008gp} with a jet radius $R=0.4$.
\begin{table}[!htbp]
\begin{center}
\begin{tabular}{c|c}
  \hline\hline
  Fiducial selection requirement  & Cut value \\
  \hline\hline
  $p_T^{\ell}$ & $>25\,\mathrm{GeV}$ \\
  $|y_{\ell}|$ & $<2.5$ \\
  $M_{e\mu}$ & $>10\,\mathrm{GeV}$ \\
  Number of jets with $p_T> 30\,\mathrm{GeV}$ & 0 \\
  $\ETmissrel$ & $>15\,\mathrm{GeV}$ \\
  $\ETmiss$ & $>20\,\mathrm{GeV}$ \\
  \hline\hline
\end{tabular}
 \end{center}
 \caption{
   Definition of the $WW\rightarrow e\mu$ fiducial phase space, where
   $p_T^\ell,y_\ell$ are the transverse momentum and rapidity of either an
   electron or a muon, $M_{e\mu}$ is the invariant mass of the electron-muon
   pair, $\ETmiss$ is the missing transverse energy, and $\ETmissrel$ is defined
   in eq.~\eqref{eq:etmissrel}.
 }
\label{tab:cuts}
\end{table}
In table~\ref{tab:cuts} we encounter the newly introduced observable
$\ETmissrel$, which is defined as follows~\cite{ATLAS:2014xea}:
\begin{equation}
  \label{eq:etmissrel}
  \ETmissrel =
    \begin{cases}
      \ETmiss \sin \Delta\phi \,, &\quad \text{if } \Delta\phi \le \frac{\pi}{2} \\
      \ETmiss \,, &\quad \text{if } \Delta\phi > \frac{\pi}{2}
    \end{cases}
\end{equation}
with
$\Delta\phi = \min\left(|\phi_e - \phi_\text{MET}|,\, |\phi_\mu -
  \phi_\text{MET}|\right)$, and $\phi_e$, $\phi_\mu$ and
$\phi_\text{MET}$ the azimuthal angle of the electron, the muon and
the missing transverse energy respectively. 

In our analysis, we omit $b$ quark-initiated contributions to
$pp\to WW$.  At LO, the $b\bar{b}$ scattering subprocess contributes
only $1\%$ to the cross section.  The $gb$ and $g\bar{b}$
subprocesses, which enter at NLO QCD increase the NLO cross section by
a factor $1.5$.  This large increase is due to graphs like
$gb\to W^- (t\to W^+b)$.  Such graphs feature a resonant top quark
propagator, which effects an enhancement of
$\mathcal{O}(m_t/\Gamma_t)=\mathcal{O}(10^2)$, which compensates the
$\mathcal{O}(1\%)$ suppression due to the $b$ PDF, and altogether an
$\mathcal{O}(1)$ contribution is obtained.  This contribution is
commonly attributed to $Wt$ production and decay (at LO QCD)
\cite{Campbell:2005bb}, and hence has to be omitted in the NLO QCD
corrections to $WW$ production, which we consider here.

We now produce both NLO, NNLL resummed, and
matched NLO+NNLL (with the matching procedure explained in
appendix~\ref{sec:th-uncerts}) predictions for the differential
distribution $d\sigma/dM_{WW}$ using \pdfforlhc\ parton distribution
functions (PDFs) at NLO~\cite{Butterworth:2015oua}, accessed through
\texttt{LHAPDF6}~\cite{Buckley:2014ana}, corresponding to
$\alpha_s(M_Z) = 0.118$, and we set both renormalisation and
factorisation scales at $M_{WW}/2$, as customary in Higgs precision
studies~\cite{Dittmaier:2011ti}.
\begin{figure}[htbp]
  \centering
  \includegraphics[width=.7\textwidth]{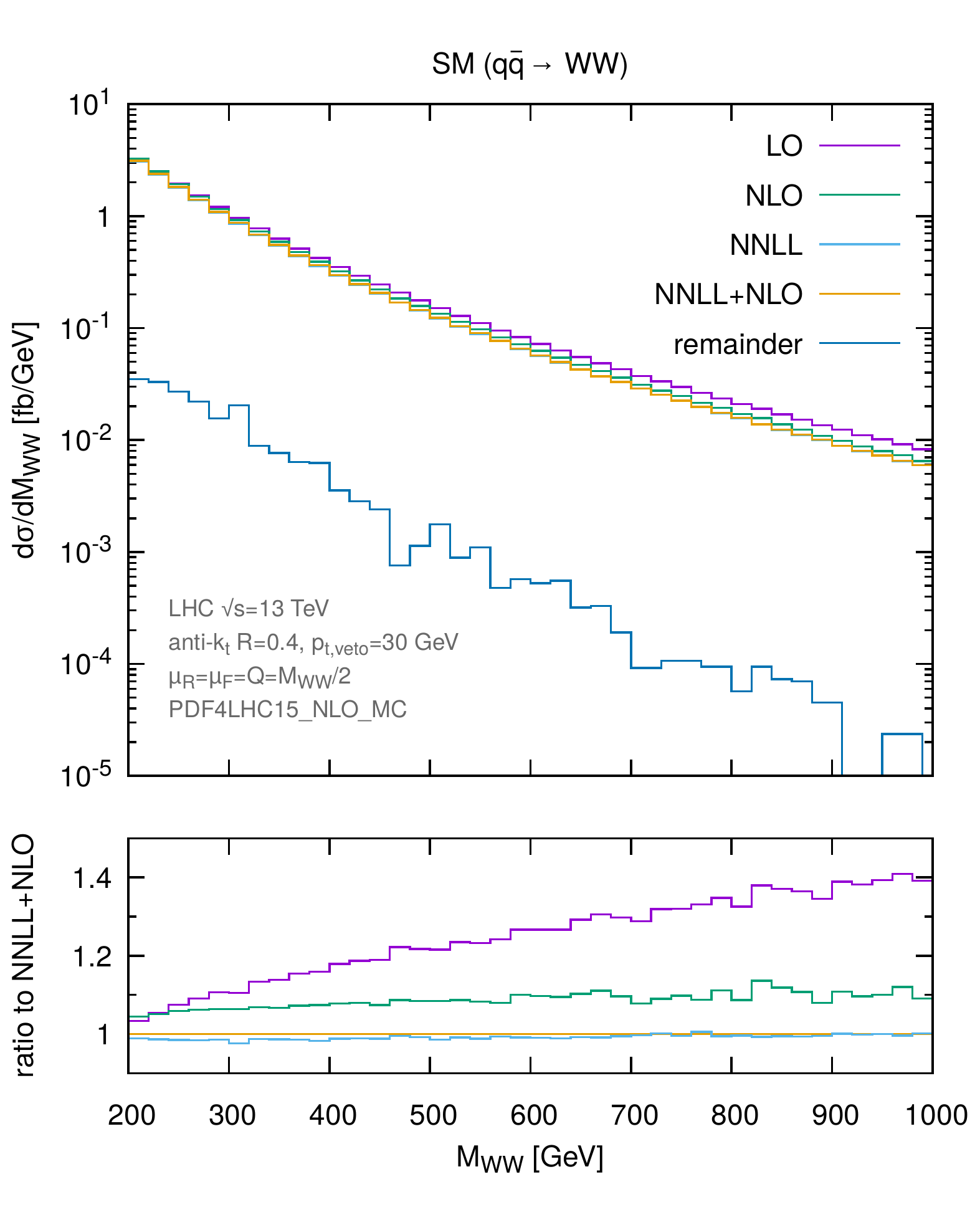}
  \caption{The differential distribution $d\sigma/dM_{WW}$ in the
    Standard Model, computed at different accuracies, and for the cuts
    described in the main text.}
  \label{fig:nlo-nnll}
\end{figure}
Fig.~\ref{fig:nlo-nnll} shows the differential cross section in the
invariant mass $M_{WW}$ of the $WW$ pair. We first note that both NLO
and NNLL+NLO are both smaller than the LO, as expected due to the
presence of a jet veto, with the suppression with respect to LO
increasing with $M_{WW}$. This implies that, in this situation, a
naive Born-level calculation fails to capture this effect and that, in
the absence of a resummation, one should use at least a NLO
prediction. NNLL+NLO gives a mild extra suppression with respect to
NLO, revealing that logarithms are not particularly large in the
considered kinematical region. However, we note that the difference
between pure NNLL resummed and matched NNLL+NLO (the so-called
``remainder''), which contains the part of the NLO which is not
enhanced by logarithms, is basically negligible. This means that the
resummation alone is very close to the best prediction we have at this
order. This is remarkable in view of the fact that to obtain NNLL
predictions we need to perform a calculation with Born-level
kinematics. On the contrary, the computational cost of the NLO
calculation is larger due to the presence of an extra emission,
without any significant gain in accuracy compared to the NNLL
prediction.

\begin{figure}[htbp]
  \centering
  \includegraphics[width=.6\textwidth]{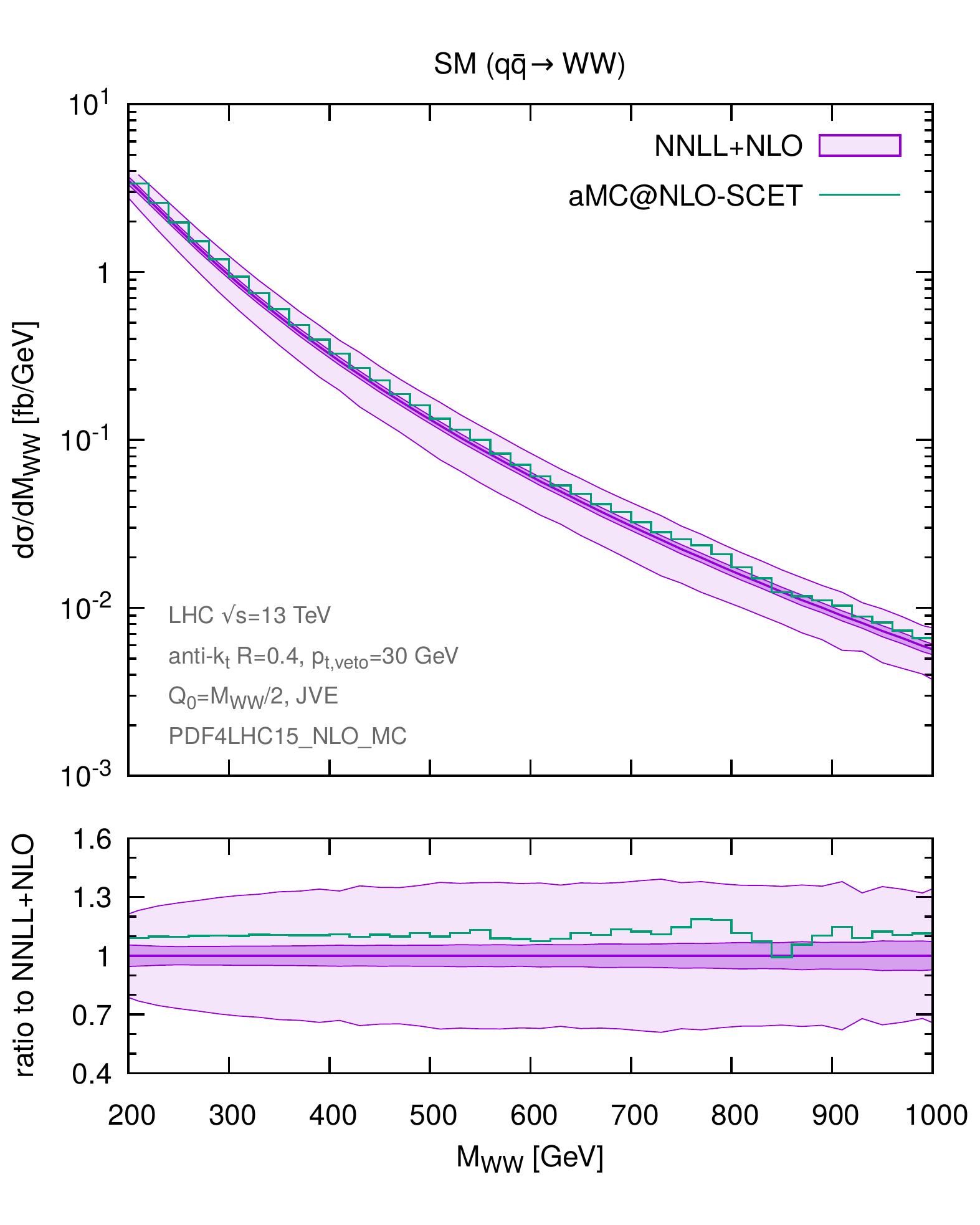}

  \vspace{-0.5cm}
  
  \caption{The differential distribution $d\sigma/dM_{WW}$ in the
    Standard Model, computed with our method, and with the program
    {\tt aMC@NLO-SCET}~\cite{Becher:2014aya}. See the main text for
    details.}
  \label{fig:scet-nnll}
\end{figure}

To complete our discussion of the $q\bar q$ channel, we compare our
predictions to those obtained from SCET via the program \texttt{aMC@NLO-SCET} of ref.~\cite{Becher:2014aya}. The comparison is shown
in Fig.~\ref{fig:scet-nnll}.
Our results contains theoretical uncertainties evaluated both with the
most recent jet-veto efficiency (JVE) method~\cite{Banfi:2015pju} at
the relevant accuracy (the wider, lighter band), and pure scale
variations (the tighter, darker band). The details of both
prescriptions can be found in appendix~\ref{sec:formulae}. The SCET
prediction corresponds to the default scale choices, and is well
within JVE uncertainties, but slightly outside the boundary of scale
variation uncertainties. Note that the most recent JVE
prescription~\cite{Banfi:2015pju} requires the so-called ``resummation
scale'' (the scale up to which soft-collinear resummations are assumed
to be valid) to be varied by a factor of 1.5 rather than the factor of 2 used to vary
renormalisation and factorisation scales (see
appendix~\ref{sec:formulae} for details). We have checked that
also varying the resummation scale by a factor of 2 does not
significantly increase the scale-uncertainty band, and the central
prediction of~\cite{Becher:2014aya} still lies outside that band. We
remark that we do not expect perfect agreement, because, although our
methods and that of~\cite{Becher:2014aya} share the same formal
accuracy, they differ in the treatment of subleading effects. Our
analysis suggests that scale uncertainties are not sufficient to
capture the size of the neglected effects, and that other methods,
such as the JVE or the one of~\cite{Boughezal:2013oha}, should be
employed to have a more realistic estimate of theoretical
uncertainties. A last comment is in order here. Within \mcfm, we do
not have access to NNLO calculations for di-boson production, so we
cannot match our resummed predictions to NNLO. As a result of this,
the JVE method may be overly conservative, due to the largish
($\sim 1.5$) K-factor of the $WW$ inclusive total cross-section, which
propagates in the evaluation of the uncertainty according to the JVE
method. If we could match to NNLO, the JVE uncertainty would be
reduced and, as happens for Higgs production~\cite{Banfi:2012jm},
would probably get closer to plain scale uncertainties.

\begin{figure}[htbp]
  \begin{minipage}[l]{0.5\linewidth}
    \begin{center}
      \includegraphics[width=\textwidth]{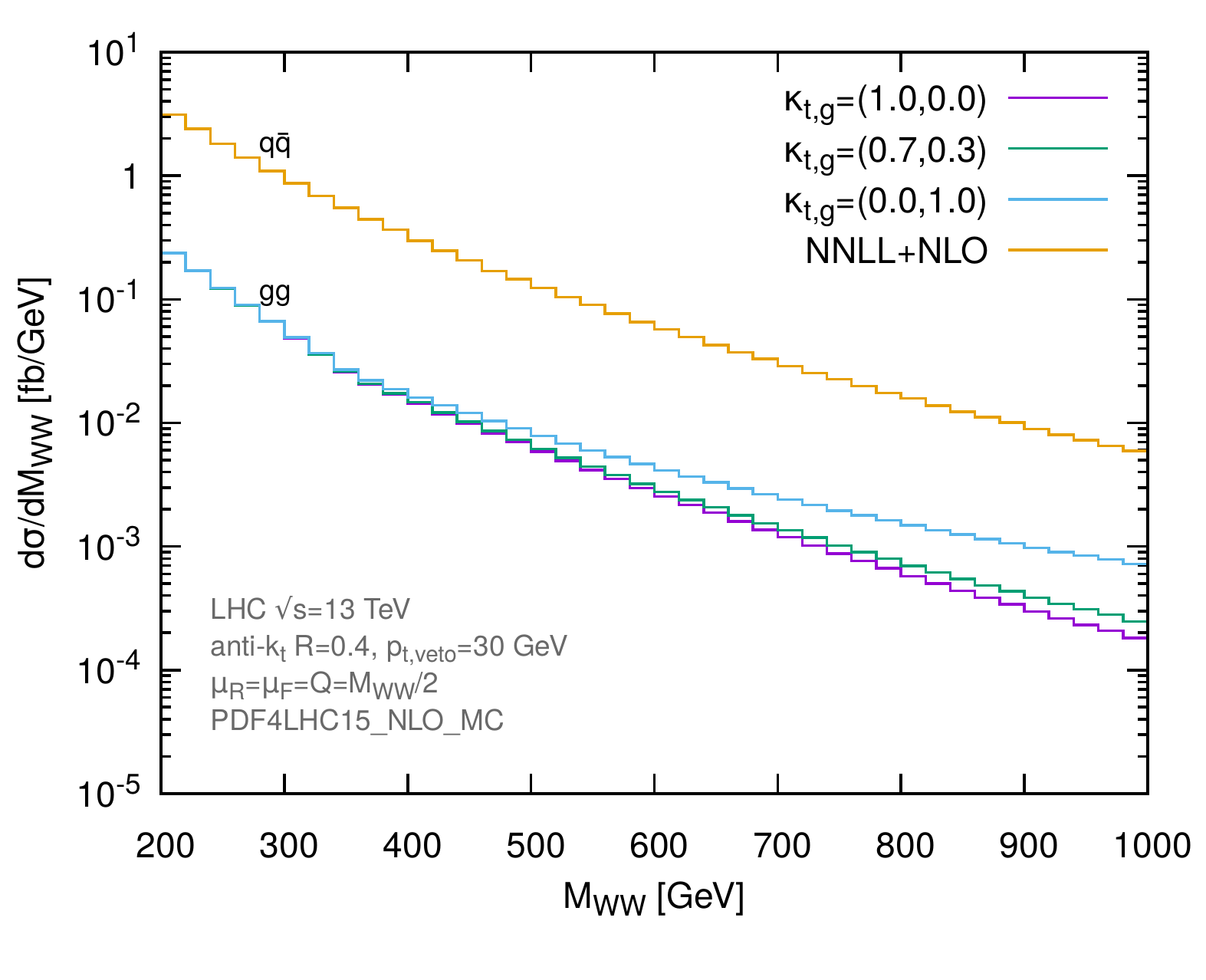}      
    \end{center}
  \end{minipage}
  \begin{minipage}[r]{0.5\linewidth}
    \begin{center}
      \includegraphics[width=\textwidth]{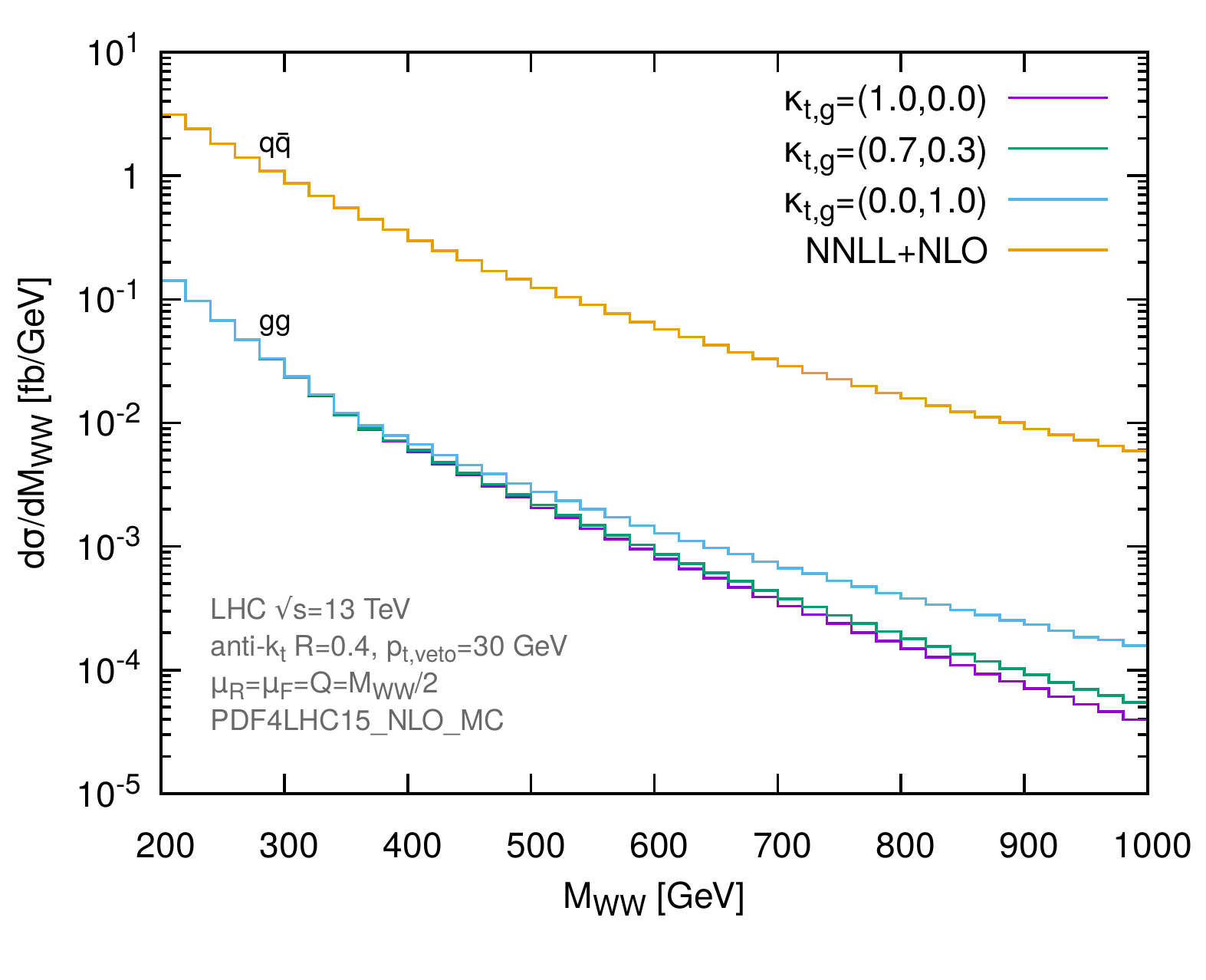}      
    \end{center}
  \end{minipage}
  \caption{The differential cross section $d\sigma/dM_{WW}$ for the three
    benchmark scenarios of eq.~\eqref{eq:benchmarks}, at LO (left) and at NLL
    (right) accuracy.}
  \label{fig:nll-lo}
\end{figure}

\begin{figure}[htbp]
  \begin{minipage}[l]{0.5\linewidth}
    \begin{center}
      \includegraphics[width=\textwidth]{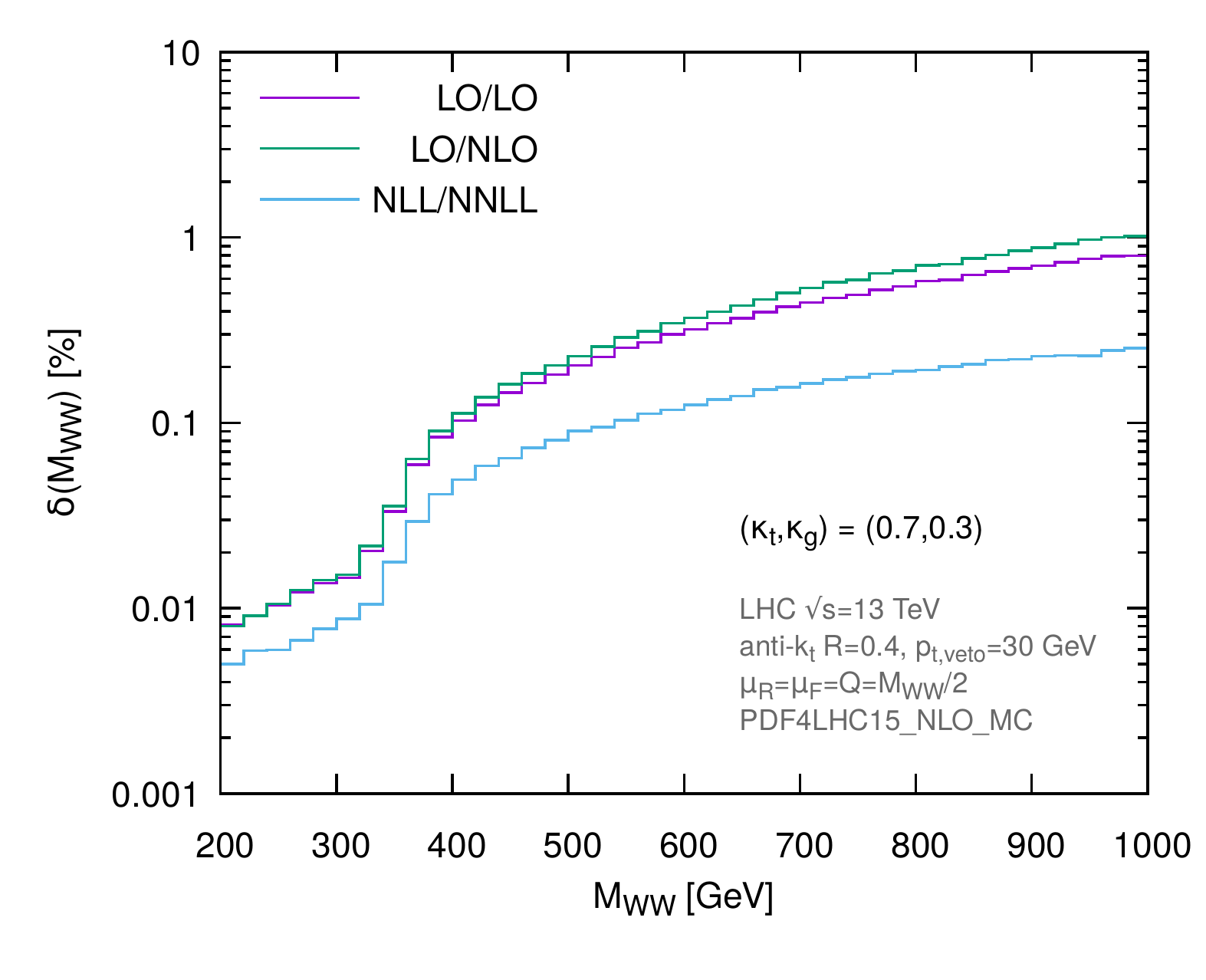}
    \end{center}
  \end{minipage}
  \begin{minipage}[r]{0.5\linewidth}
    \begin{center}
      \includegraphics[width=\textwidth]{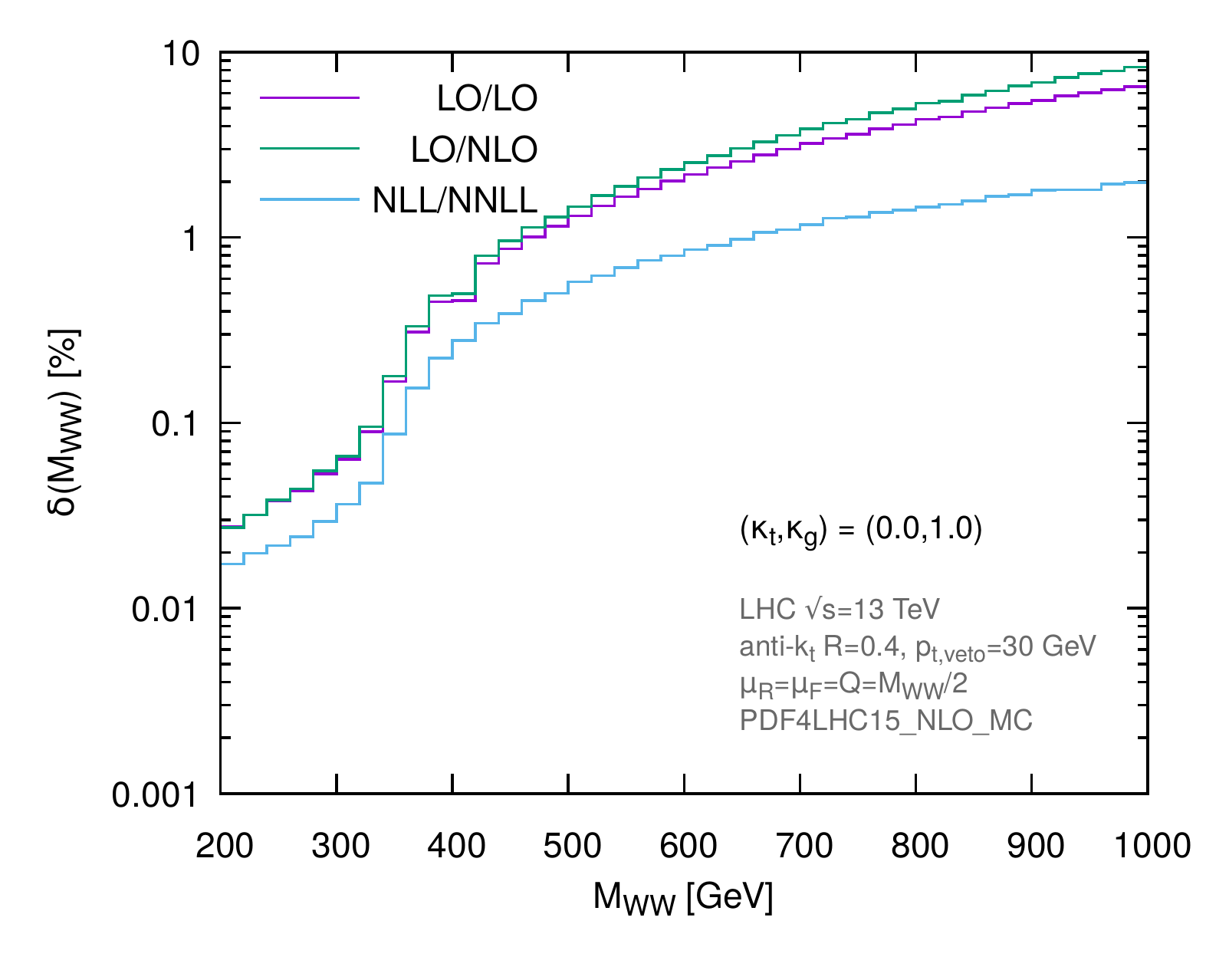}
    \end{center}
  \end{minipage}
  \caption{The relative difference between BSM and SM
    $d\sigma/dM_{WW}$ defined in eq.~(\ref{eq:delta}) for the two
    benchmark scenarios $ (\kappa_t,\kappa_g)_{\rm BSM_1}$ (left) and
    $ (\kappa_t,\kappa_g)_{\rm BSM_2}$ (right). The labels refer to
    the accuracy employed in the calculation of numerator and
    denominator in eq.~(\ref{eq:delta}).}
  \label{fig:delta}
\end{figure}

In order to have a specific example of a BSM theory that implements
the effective operator of eq.~(\ref{eq:BSM-operator}), we consider the
following modification of the SM Lagrangian~\cite{Grojean:2013nya}:
\begin{equation}
  \label{eq:BSM-model}
  \mathcal{L} \supset -\kappa_t \frac{m_t}{v} h \bar t t + \kappa_g\frac{\alpha_s}{12\pi}\frac{h}{v} G_{\mu\nu}^a G^{\mu\nu}_a\,,
\end{equation}
with $t$, $h$, $G_{\mu\nu}^a$ the top field, the SM Higgs field, and
the gluon field strength respectively. The SM corresponds to
$(\kappa_t,\kappa_g)=(1,0)$, and in this section we will only explore
BSM scenarios such that $\kappa_t+\kappa_g=1$, which ensures that the
Higgs total cross section stays unchanged (modulo quark-mass effects,
which give a correction of a few percent~\cite{Banfi:2015pju}). Such
modifications of the SM Lagrangian only affect the gluon-fusion
contribution to di-boson production. Their effect has been
investigated before for the case of $ZZ$
production~\cite{Buschmann:2014sia}, where one does not need to impose
a jet veto to suppress unwanted background. Here we wish to study how
the presence of a jet veto, required for studies of $WW$ production,
affects the relative size of a BSM contribution with respect to the SM
background. We consider the three benchmark scenarios studied in
ref.~\cite{Buschmann:2014sia}, i.e.\
\begin{equation}
  \label{eq:benchmarks}
(\kappa_t,\kappa_g)_{\rm
  SM}=(1,0)\,,\quad (\kappa_t,\kappa_g)_{\rm BSM_1}=(0.7,0.3)\,,\quad(\kappa_t,\kappa_g)_{\rm BSM_2}=(0,1)\,.
\end{equation}
First, in Fig.~\ref{fig:nll-lo} we compare the loop-induced gluon
fusion contribution to the $M_{WW}$ distribution at LO, which is what
is given by default by any automated Born-level event generator, with
the NLL analytic resummation, which gives the best modelling of
jet-veto effects at the currently available accuracy. Our best
$q\bar q$ prediction is also shown for comparison.
We see that, if we include resummation effects, the cross section for
each benchmark point is reduced by almost an order of magnitude in the
tail of the distribution, where BSM effects start to become
important. We then investigate more quantitatively how this impacts
the deviations we might observe with respect to the SM, by plotting
the quantity
\begin{equation}
  \label{eq:delta}
  \delta(M_{WW}) = \frac{d\sigma^{\rm BSM}_{gg}/dM_{WW}-d\sigma^{\rm SM}_{gg}/dM_{WW}}{d\sigma_{q\bar q}/dM_{WW}}\,.
\end{equation}
In the above equation, $d\sigma^{\rm BSM}_{gg}$ is computed according
to eq.~(\ref{eq:dsigma-BSM}), $d\sigma^{\rm SM}_{gg}$ differs from
$d\sigma^{\rm BSM}_{gg}$ by the fact that the BSM contribution to the
amplitude ($M_{gg}^{\rm BSM}$ in eq.~(\ref{eq:LBSM})) is set to zero,
and $d\sigma_{q\bar q}$ follows from
eq.~(\ref{eq:fullxsc}). Fig.~\ref{fig:delta} (left) shows
$\delta(M_{WW})$ for the benchmark point $(0.7,0.3)$. We first note
the growth of this quantity with energy, as expected from the
effective nature of the $ggH$ vertex. Fortunately, the growth persists
after including jet-veto effects through NLL resummation, however the
deviation from the SM reduces from the 1\% that one would obtain using
fixed-order calculations (see fig.~\ref{fig:nll-lo}) to fractions of a
percent. The same quantity shown in the right panel of
fig.~\ref{fig:delta} for the benchmark point $(0.0,1.0)$ displays
qualitatively the same behaviour, although the deviation is a factor
ten bigger. We see that, in the presence of jet-veto restrictions such
as the one in ATLAS cuts~\cite{Aaboud:2017qkn}, one is bound to use a
theoretical tool that resums large logarithms. This could be either
resummed predictions, or simulations with parton-shower event
generators.

\begin{figure}[htbp]
  \begin{minipage}[l]{0.5\linewidth}
    \begin{center}
      \includegraphics[width=\textwidth]{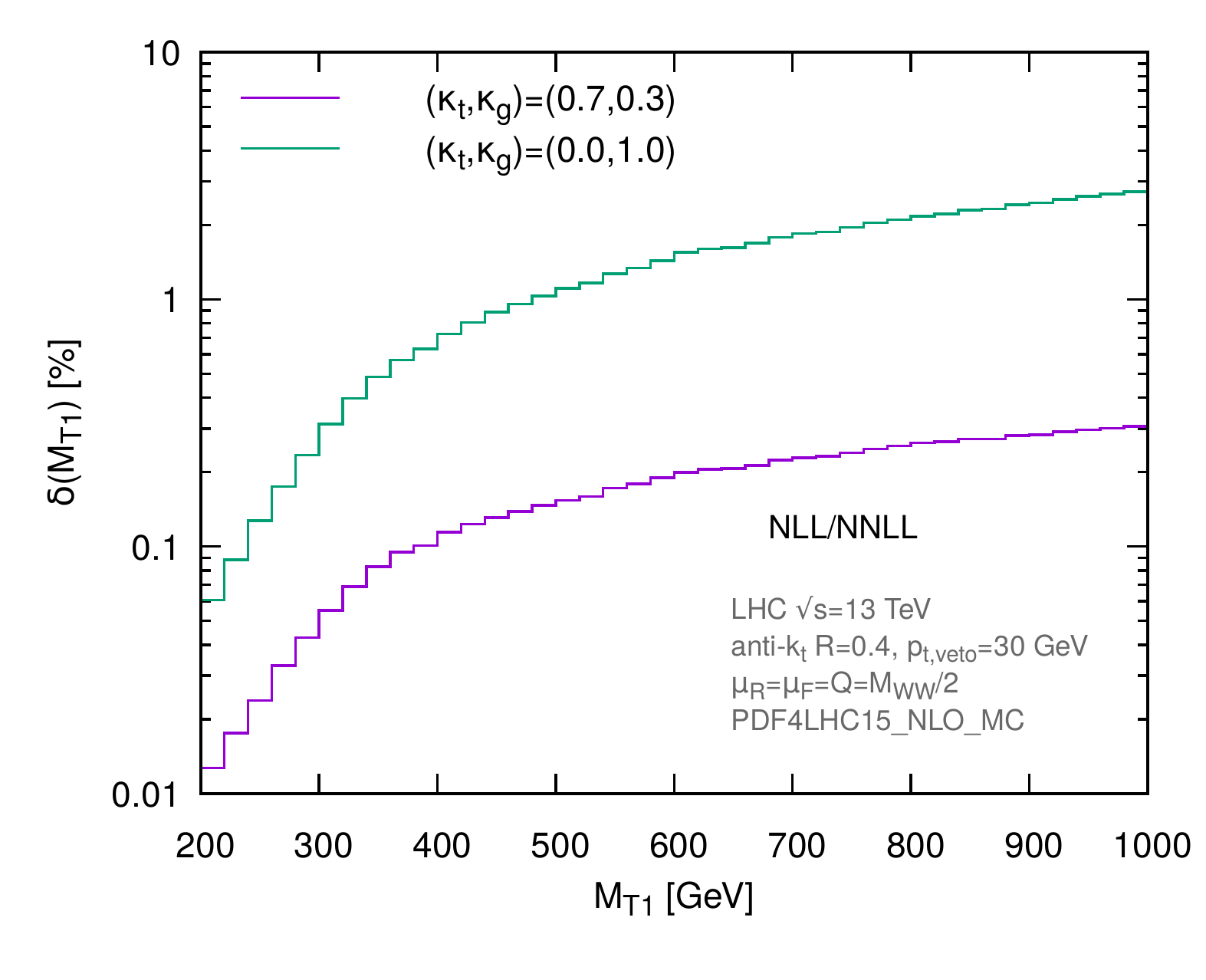}
    \end{center}
  \end{minipage}
  \begin{minipage}[r]{0.5\linewidth}
    \begin{center}
      \includegraphics[width=\textwidth]{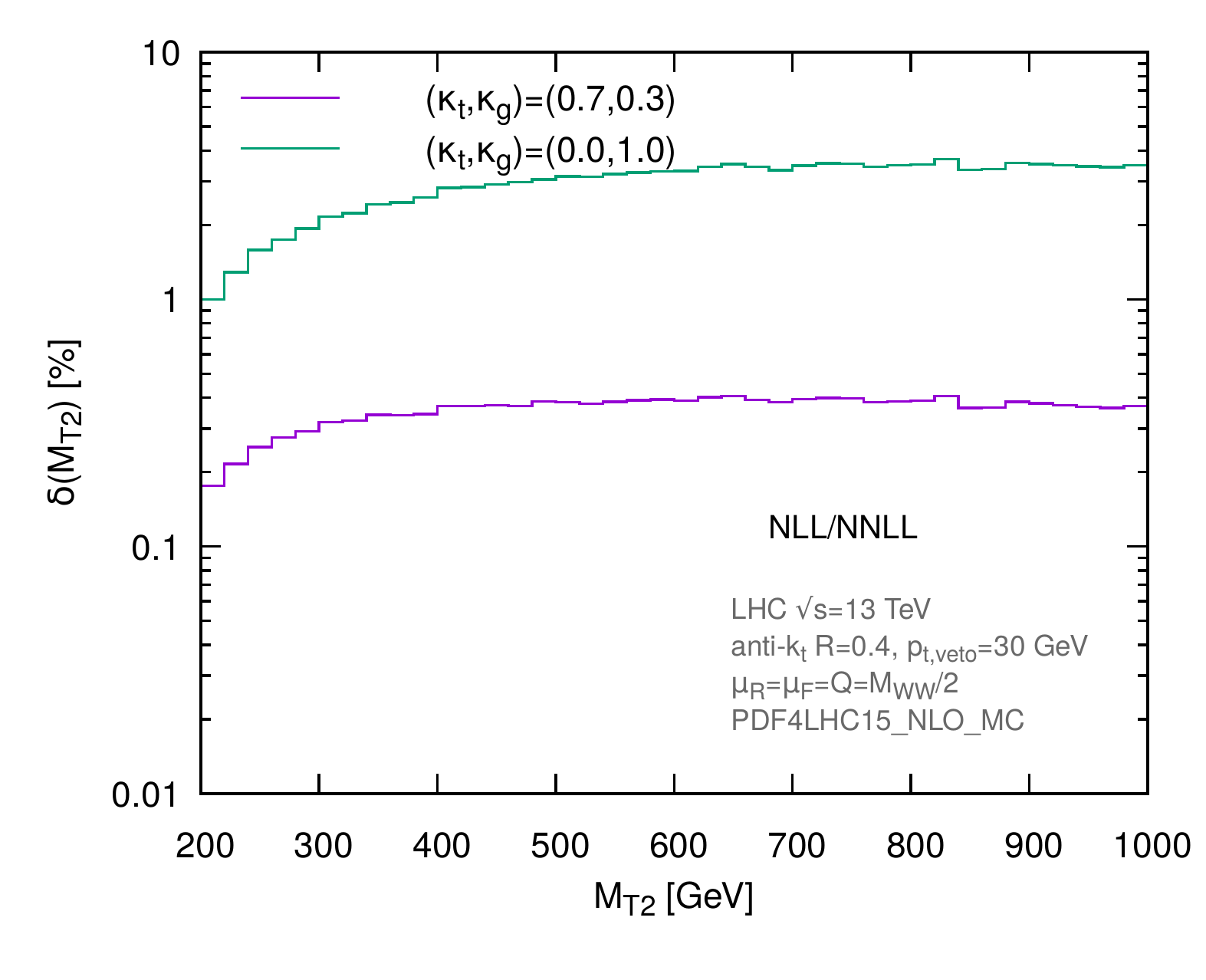}
    \end{center}
  \end{minipage}
  \caption{The relative difference between BSM and Standard Model $WW$
    production, differential in $M_{T1}$ (left) and $M_{T2}$ (right).}
  \label{fig:delta-MT12}
\end{figure}

\begin{figure}[htbp]
  \centering
  \includegraphics[width=.7\textwidth]{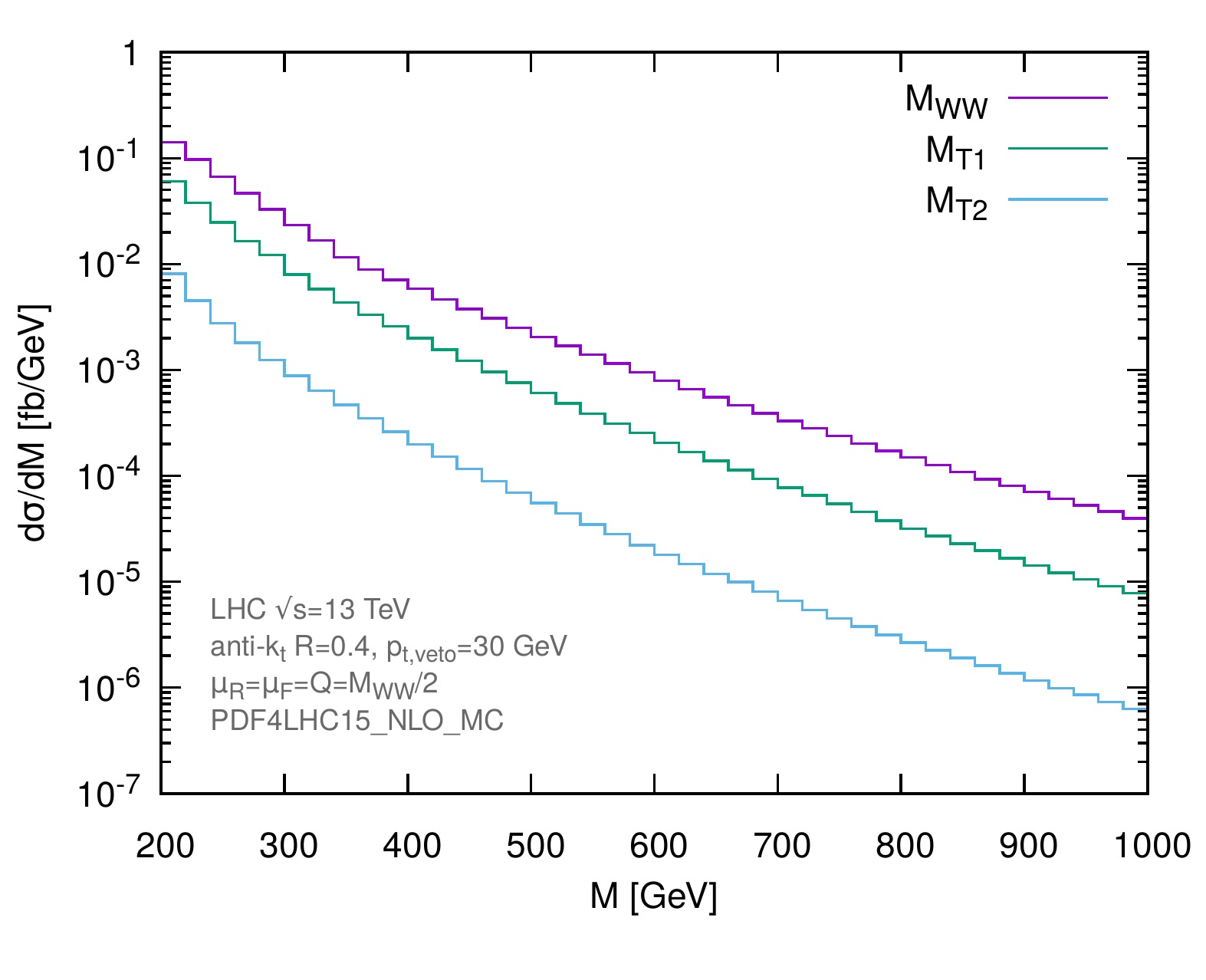}
  \vspace{-.5cm}
  
  \caption{The distributions in $M=M_{WW},M_{T1},M_{T2}$ for the $gg$
    incoming channel.}
  \label{fig:SM-MT1-MT2}
\end{figure}

The variable $\delta(M_{WW})$ is of theoretical interest only, because
we do not have access to the momenta of the neutrinos. To have
experimentally accessible observables, we consider differential
distributions in $M_{T1}$~\cite{Aad:2012uub},
$M_{T2}$~\cite{Chatrchyan:2012ty} and $M_{T3}$~\cite{Aad:2012uub},
three measurable variables that are strongly correlated with $M_{WW}$
\begin{subequations}
  \begin{align}
      \label{eq:MT1}
  M_{T1} & = \sqrt{(M_{\rm{T},e\mu} + {p\!\!\!/}_{\rm T})^2 - (\vec{p}_{\rm{T},e\mu} + \vec{p\!\!\!/}_{\rm T})^2} \,,
  \quad M_{\rm{T},e\mu} = \sqrt{p_{\rm{T}, e\mu}^2 + M_{e\mu}^2}\,, \\
  \label{eq:MT2}
  M_{T2} & = \sqrt{2 p_{\rm{T},e\mu} {p\!\!\!/}_{\rm T}(1 - \cos \Delta\phi_{e\mu,\rm{miss}})}\,, \\
  \label{eq:MT3}
  M_{T3} & = \sqrt{(M_{\rm{T},e\mu} + {M\!\!\!\!/}_{\rm T})^2 - (\vec{p}_{\rm{T},e\mu} + \vec{p\!\!\!/}_{\rm T})^2}\,, 
  \quad {M\!\!\!\!/}_{\rm T} = \sqrt{{p\!\!\!/}_{\rm T}^2 + M_{e\mu}^2}\,.
  \end{align}
\end{subequations}
In the above equations,
$\vec{p}_{\rm{T},e\mu}=\vec p_{T,e}+\vec p_{T,\mu}$, and
$ M_{e\mu}^2=(p_e+p_\mu)^2$. The vector $\vec{p\!\!\!/}_{\rm T}$ is
the missing transverse momentum, defined as minus the vector sum of
all detectable particles. Note that, if no jets are present, as at
Born-level and in NNLL resummed predictions,
$\vec{p\!\!\!/}_{\rm T}=-\vec{p}_{\rm{T},e\mu}$. Last,
$\Delta\phi_{e\mu,\rm{miss}}$ is the azimuthal angle between
$\vec{p}_{\rm{T},e\mu}$ and $\vec{p\!\!\!/}_{\rm T}$.
The corresponding results for $\delta$ are shown in
Fig.~\ref{fig:delta-MT12}.
We note that $M_{T2}$ gives rise to considerably larger deviations
with respect to $M_{T1}$. This is because low values of $M_{T2}$ are
correlated to larger values of $M_{WW}$, so $M_{T2}$ effectively
probes the $M_{WW}$ distribution in the high-mass tail, where BSM
effects are appreciable. However, this also means that the
differential cross section in $M_{T2}$ is much smaller than that in
$M_{T1}$, as can be seen from Fig.~\ref{fig:SM-MT1-MT2}.
Therefore, the discriminatory power of $M_{T2}$ is only of use if
we have a very large number of events. We have also studied the variable $M_{T3}$ defined again in
ref.~\cite{Chatrchyan:2012ty} and first devised in
ref.~\cite{Rainwater:1999sd}.  The distribution in this variable looks
very similar to that of $2 M_{T1}$, so the same discussion as for
$M_{T1}$ applies here.

We now compare our results to parton-level predictions from
parton-shower event generators, using existing tunes. In particular,
for $q\bar q$ we consider
\powheg~\cite{Alioli:2010xd,Frixione:2007vw,Nason:2004rx,Nason:2013ydw}
matched to the \aznlo~\cite{Aad:2015auj} tune of
\pythiaeight~\cite{Sjostrand:2014zea}, and
\amcatnlo~\cite{Alwall:2014hca,Frixione:2010ra,Frixione:2002ik,Frixione:1997np,Frixione:1995ms}
matched to \pythia, this time with the default parameters. To
investigate the dependence on the shower algorithm, we also consider
the parton shower \herwigseven~\cite{Bahr:2008pv,Bellm:2015jjp}
matched as \powherwig, and \mcherwig, both with the default
parameters. For \powpythia, we use the PDF set by the \aznlo\ tune,
i.e.\ \ctten~\cite{Lai:2010vv} for \powheg\ and
\ctsix~\cite{Pumplin:2002vw} for the parton shower. For consistency,
we use \ctten\ everywhere for \powherwig. For \powherwig, we also
performed runs with default shower PDFs, and noted no significance
difference in the resulting distributions. For all the \amcatnlo\ runs
we use \pdfforlhc\ PDFs, both for the generation of the hard
configurations and the shower.

\begin{figure}[htbp]
  \begin{minipage}[l]{0.5\linewidth}
    \includegraphics[width=\textwidth]{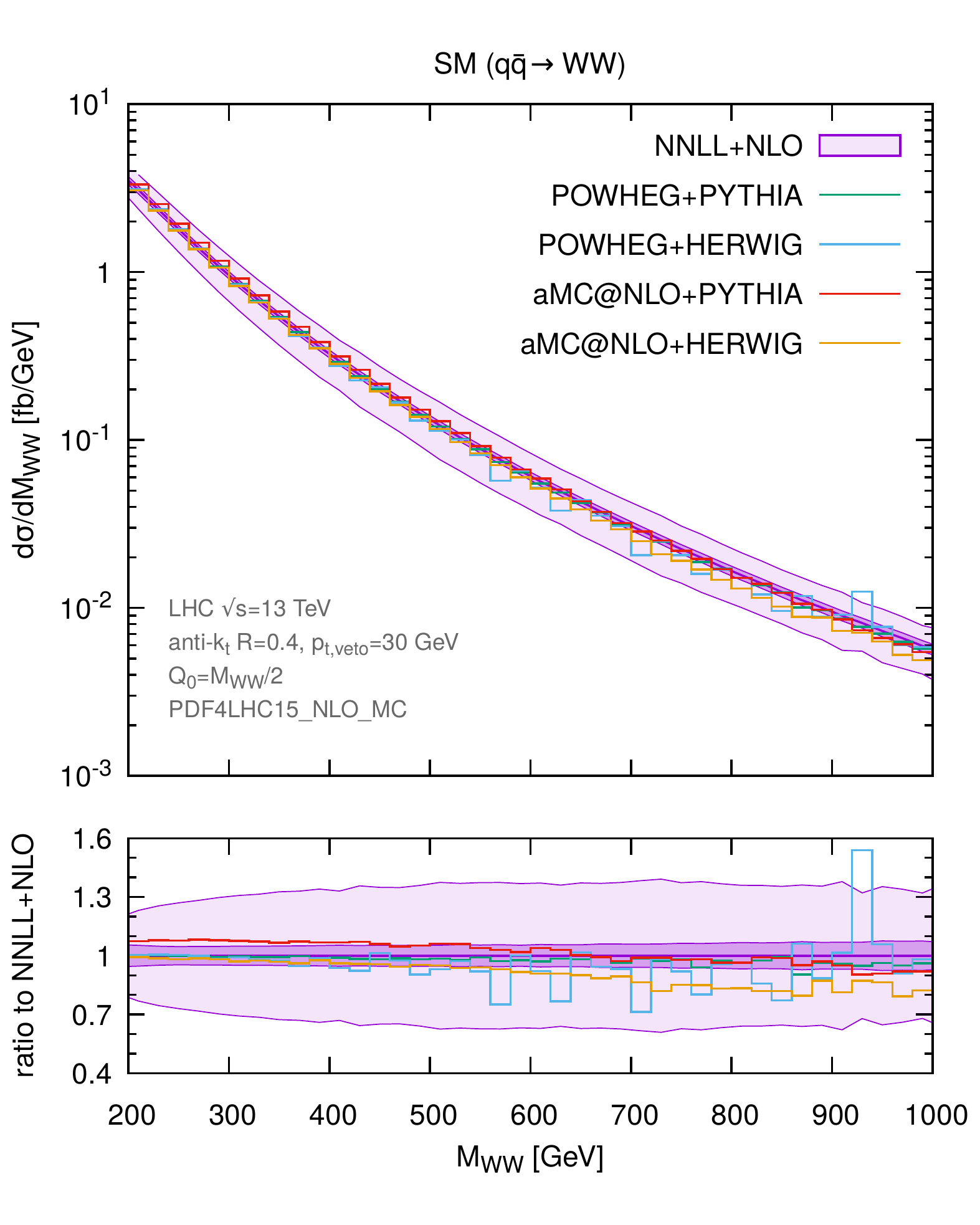}
  \end{minipage}
  \begin{minipage}[r]{0.5\linewidth}
     \includegraphics[width=\textwidth]{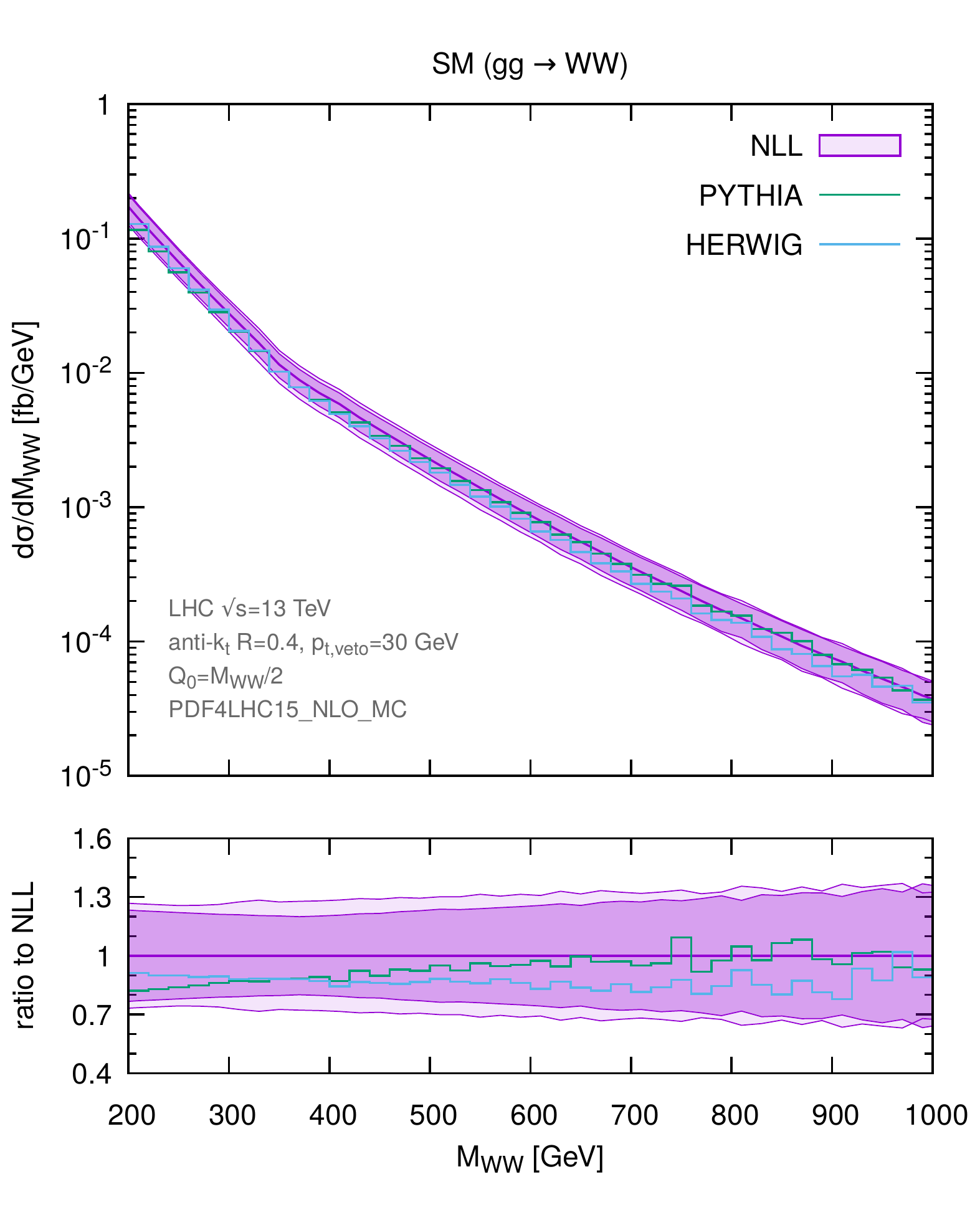}
  \end{minipage}
  \caption{Analytical predictions for the SM distribution in the
    invariant mass of a $WW$ pair, compared to results from various
    parton-shower event generators, corresponding to the details given
    in the main text.}
  \label{fig:SM-MC}
\end{figure}

\begin{figure}[htbp]
  \begin{minipage}[l]{0.5\linewidth}
    \includegraphics[width=\textwidth]{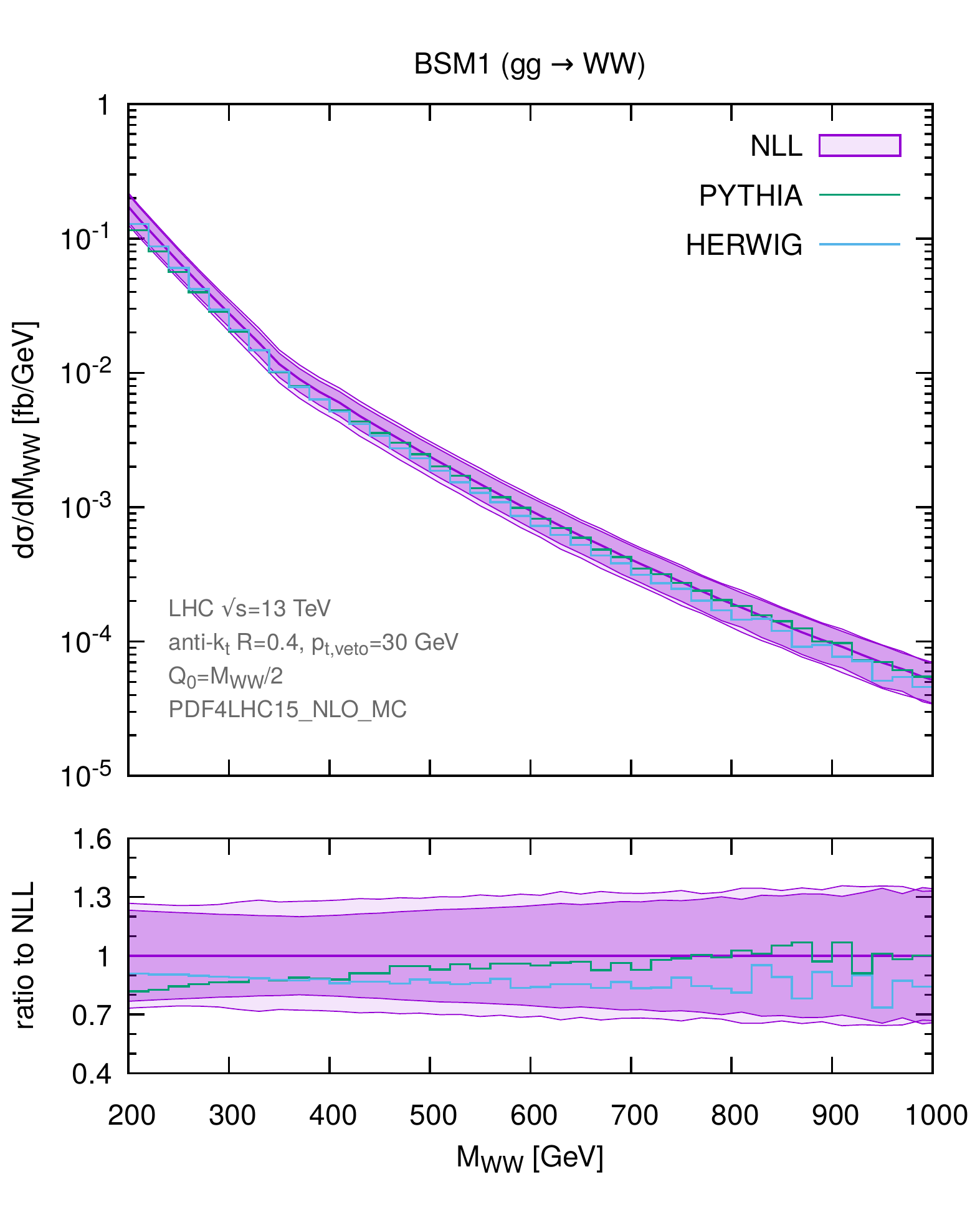}
  \end{minipage}
  \begin{minipage}[r]{0.5\linewidth}
     \includegraphics[width=\textwidth]{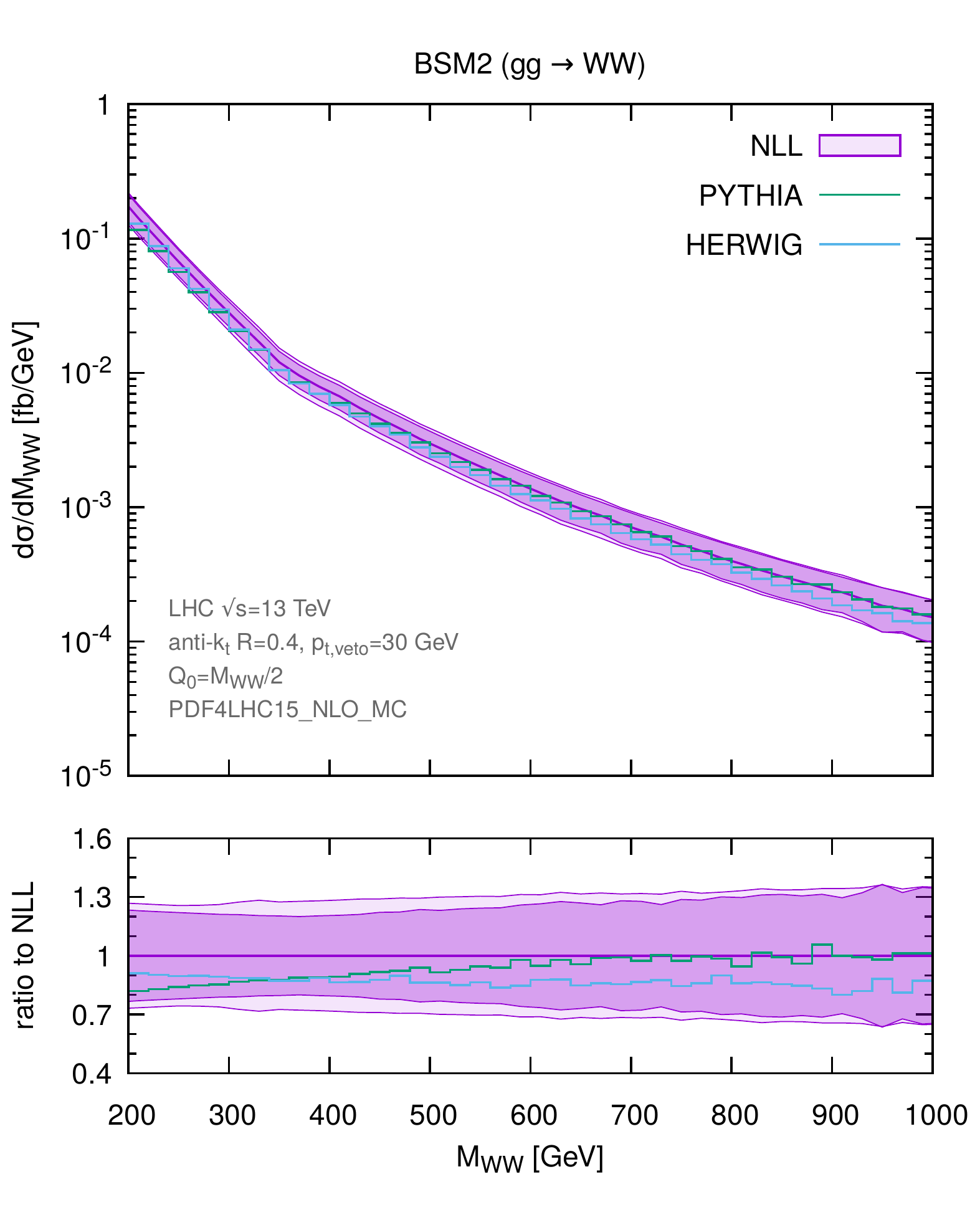}
  \end{minipage}
  \caption{The same distribution as in Fig.~\ref{fig:SM-MC} for the two BSM scenarios considered in the main text.}
  \label{fig:BSM-MC}
\end{figure}

The comparison of resummation with event generators is shown in
Fig.~\ref{fig:SM-MC} for the SM (for $q\bar q\to WW$ and $gg\to WW$
separately), and in Fig.~\ref{fig:BSM-MC} for the two BSM scenarios
considered above. Resummed predictions include an estimate of theory
uncertainties at the appropriate accuracy, as explained in
appendix~\ref{sec:th-uncerts}. Note that, due to the missing NLO total
cross-section for the incoming $gg$ channel, JVE and scale
uncertainties for $gg\to WW$ are of comparable size, with the JVE ones
slightly larger.
We first observe that, both for $q\bar q$- and for $gg$-initiated $WW$
production, all event generators agree with the resummation within its
uncertainties. For $q\bar q$, where we can match parton-shower
predictions to NLO, \powpythia\ shows a remarkable agreement with the
resummation, but other event generators give comparable results. We
note that predictions obtained with \amcatnlo\ show a slightly
different trend with $M_{WW}$. In particular \mcpythia\ is slightly
above our central prediction at low $M_{WW}$, and a bit lower at high
$M_{WW}$, whereas \mcherwig\ shows the same trend but is everywhere
lower than our predictions.

In the $gg$ case, both for the SM and the considered BSM scenarios, we
can only compare to unmatched parton-showers results, as no NLO
calculation is available.  We observe that \pythia\ is in better
agreement with our predictions at large values of $M_{WW}$, whereas
{\herwig}'s predictions have the same shape as ours, but are
systematically lower by about 10\%. Overall, there is agreement
between our predictions and parton showers within uncertainty bands,
so the latter can be reliably used for this process. We remark that
parton-shower predictions not only have lower formal accuracy, but are
also much more expensive computationally. Hence it might be lengthy to
assess with those tools if a range of BSM parameters leads to sizeable
deviations from the SM, whereas with our numerical implementation such
analyses could be performed at the cost of an unshowered Born-level
calculation.

\begin{figure}[htbp!]
  \begin{minipage}[l]{0.5\linewidth}
    \begin{center}
      \includegraphics[width=\textwidth]{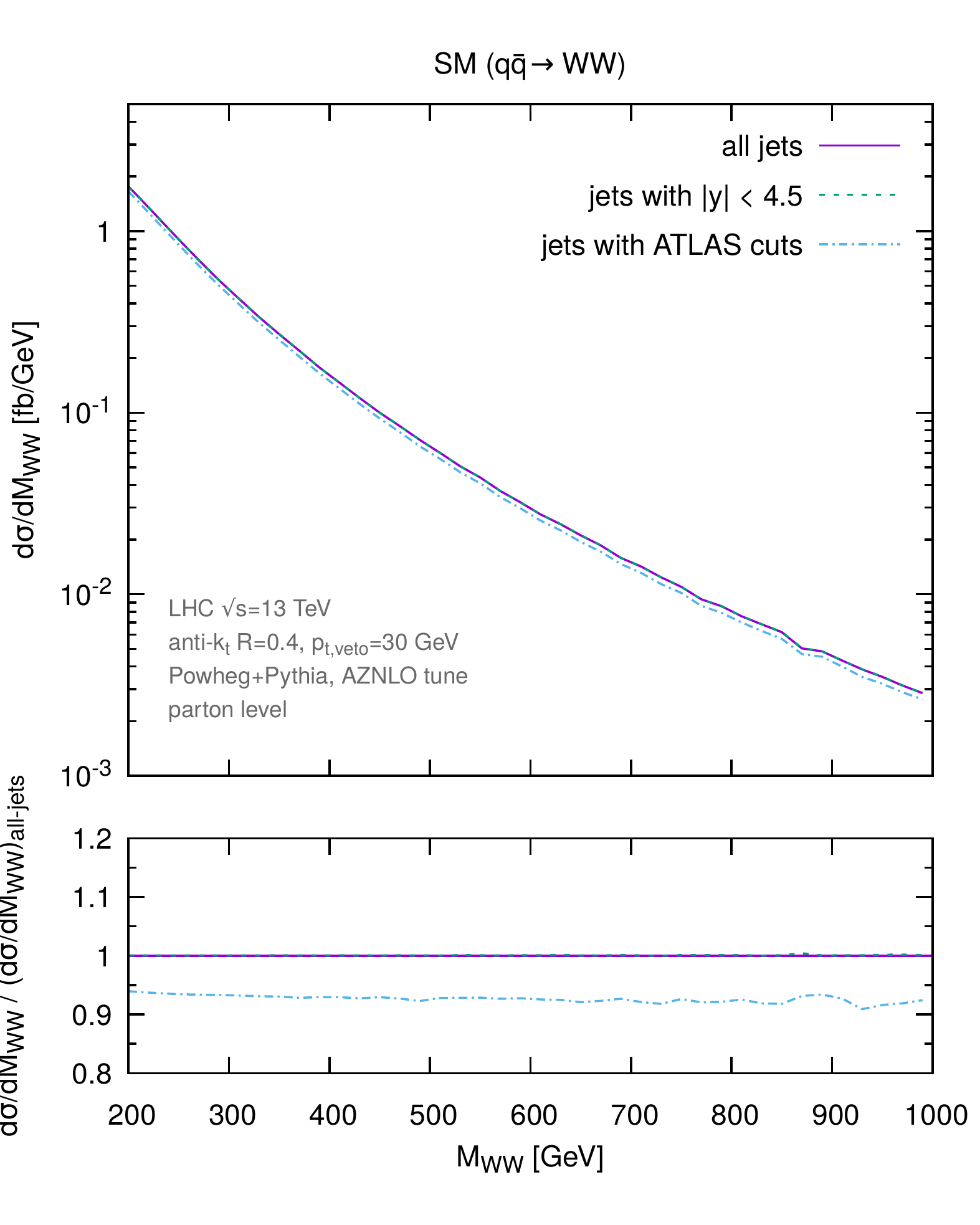}      
    \end{center}
  \end{minipage}
  \begin{minipage}[r]{0.5\linewidth}
    \begin{center}
      \includegraphics[width=\textwidth]{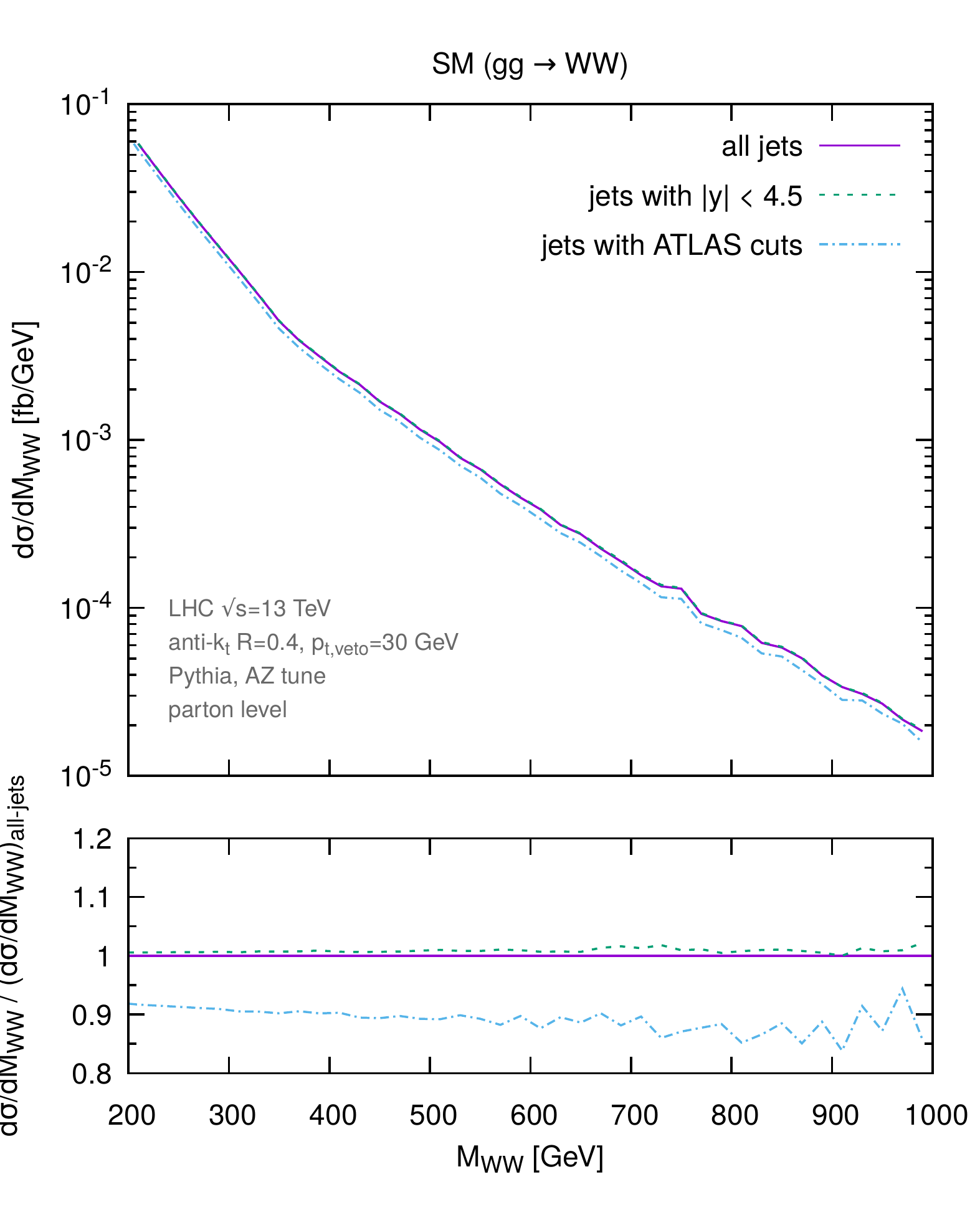}      
    \end{center}
  \end{minipage}
  \caption{Impact of different cuts on the jets on $d\sigma/dM_{WW}$ in the SM
    for $q\bar q$ (left) modelled with \powpythia\ and $gg$ (right)
    modelled with plain \pythia.}
  \label{fig:mc-fullcuts}
\end{figure}
\begin{figure}[htbp!]
  \begin{minipage}[l]{0.5\linewidth}
    \begin{center}
      \includegraphics[width=\textwidth]{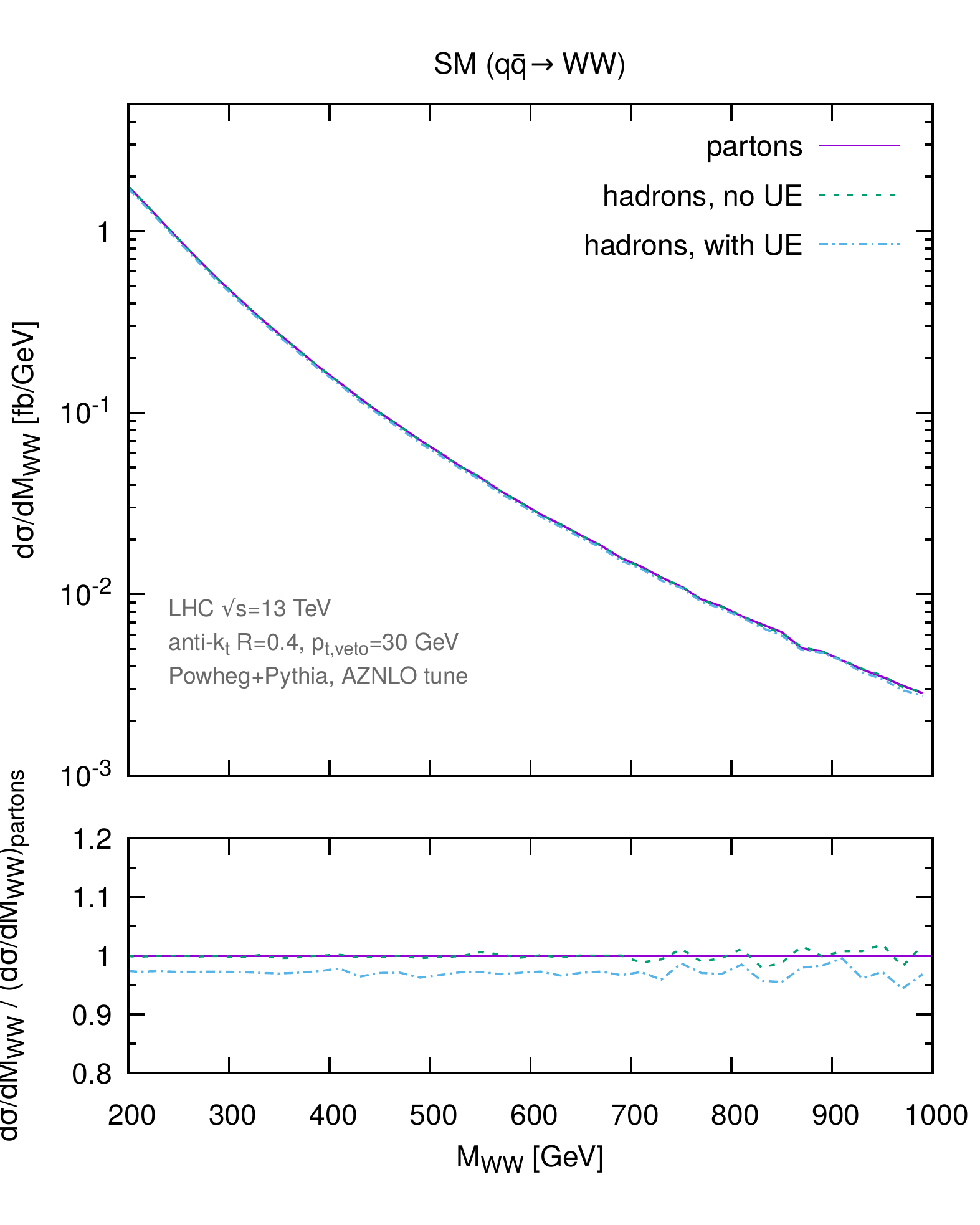}
    \end{center}
  \end{minipage}
  \begin{minipage}[r]{0.5\linewidth}
    \begin{center}
      \includegraphics[width=\textwidth]{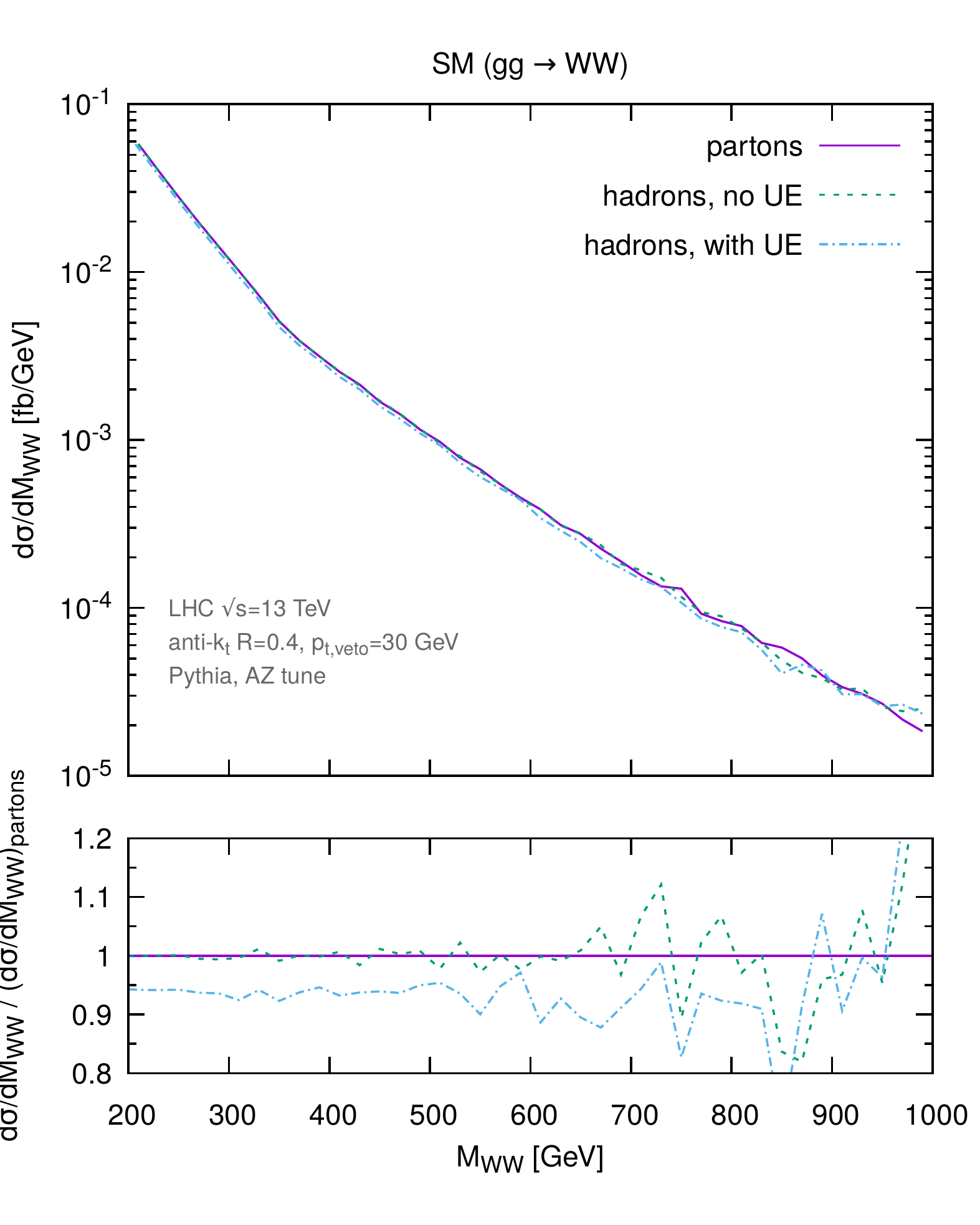}      
    \end{center}
  \end{minipage}
  \caption{Impact of hadronisation and underlying event on 
    $d\sigma/dM_{WW}$ in the SM for $q\bar q$ (left) modelled with \powpythia\
    and $gg$ (right) modelled with plain \pythia. The fluctuations in
    the right plot are due to statistical uncertainties in the Monte
    Carlo samples.}
  \label{fig:mc-np}
\end{figure}

We now investigate the impact of actual ATLAS cuts on the jets with
respect to the simplified cuts in table~\ref{tab:cuts}. First, ATLAS
vetoes only jets with $|y|<4.5$. This might cause problems for our
resummed calculation because, according to the argument of
ref.~\cite{Banfi:2004nk}, it limits its validity to
$\ln(M_{WW}/\ptjv)\lesssim 4.5$. Also, ATLAS employs an
additional cut on the jets, vetoing also jets with
$p_T>25\,\mathrm{GeV}$ and $|y|<2.5$. If we compute $d\sigma/dM_{WW}$
with the cuts in table~\ref{tab:cuts}, we miss a contribution of order
$\exp[-C (\as/\pi) \Delta y
\ln(30\,\mathrm{GeV}/25\,\mathrm{GeV})]$\footnote{This
  naive estimate neglects the so-called non-global
  logarithms~\cite{Dasgupta:2001sh}.}, with $C=C_F$ or $C=C_A$
according to whether we have quarks or gluons in the initial state and
$\Delta y$ the size of rapidity region in which the jet veto cuts
differ, in this case $\Delta y=5$. This contribution is formally NNLL,
because the rapidity region where ATLAS applies a more stringent
jet-veto cut does not increase with increasing $M_{WW}$, for fixed
$\ptjv$.  Last, the definition of $\ETmissrel$ used to define the cuts
in table~\ref{tab:cuts} considers only leptons, whereas ATLAS
considers all reconstructed particles, including jets. This leads to
small NNLL corrections that depend on the area in the $y$-$\phi$ plane
occupied by the rejected jets. We study these effects using
parton-shower event generators. In particular, in
fig.~\ref{fig:mc-fullcuts} we assess the impact of different cuts on
the jets on $d\sigma/dM_{WW}$, using parton shower event generators at
parton level, in particular we use \powpythia\ for $q\bar q$ and plain
\pythia\ for $gg$. We observe that the rapidity cut $|y|<4.5$ has
essentially no effect. On the contrary, implementing the full ATLAS
cuts gives a sizeable but constant extra suppression. This is
reasonable given that the jet veto cut imposed by ATLAS in the central
region $|y|<2.5$ is tighter than the one corresponding to our
simplified cuts. Although the contribution we miss is formally NNLL,
for the values of $M_{WW}$ we consider here, the rapidity region in
which $\ptjv = 25\,$GeV is larger than that where $\ptjv = 30\,$GeV.
Therefore, using our simplified cuts to mimic the ATLAS cuts we miss a
potentially large contribution.  In the case of $gg$, the suppression
is larger with respect to $q\bar q$ due to the larger colour factor of
the initial-state gluons with respect to the quarks.

Last, in fig.~\ref{fig:mc-np} we investigate the impact on
$d\sigma/dM_{WW}$ of non-perturbative corrections due to hadronisation
and underlying event, using parton shower event generators.  Again we
make use of \powpythia\ for $q\bar q$ and plain \pythia\ for $gg$. We
observe that hadronisation corrections are essentially negligible,
which is expected since they scale like inverse powers of the hard
scale, in this case $M_{WW}$. Corrections arising from the underlying
event are a few percent, smaller than the typical theoretical
uncertainties of our predictions.

To summarise, the effect with the greatest impact is the different
jet-veto procedure employed by ATLAS. This could be modelled more
accurately, either by making use of an effective $\ptjv$, or even
better by performing a resummation of jet-veto effects with rapidity
cuts, as done in~\cite{Michel:2018hui}. Both improvements are beyond
the scope of the present work.

\section{Sensitivity studies}
\label{sec:sensitivity}

\begin{figure}[htbp]
  \begin{minipage}[l]{0.5\linewidth}
    \begin{center}
      \includegraphics[width=\textwidth]{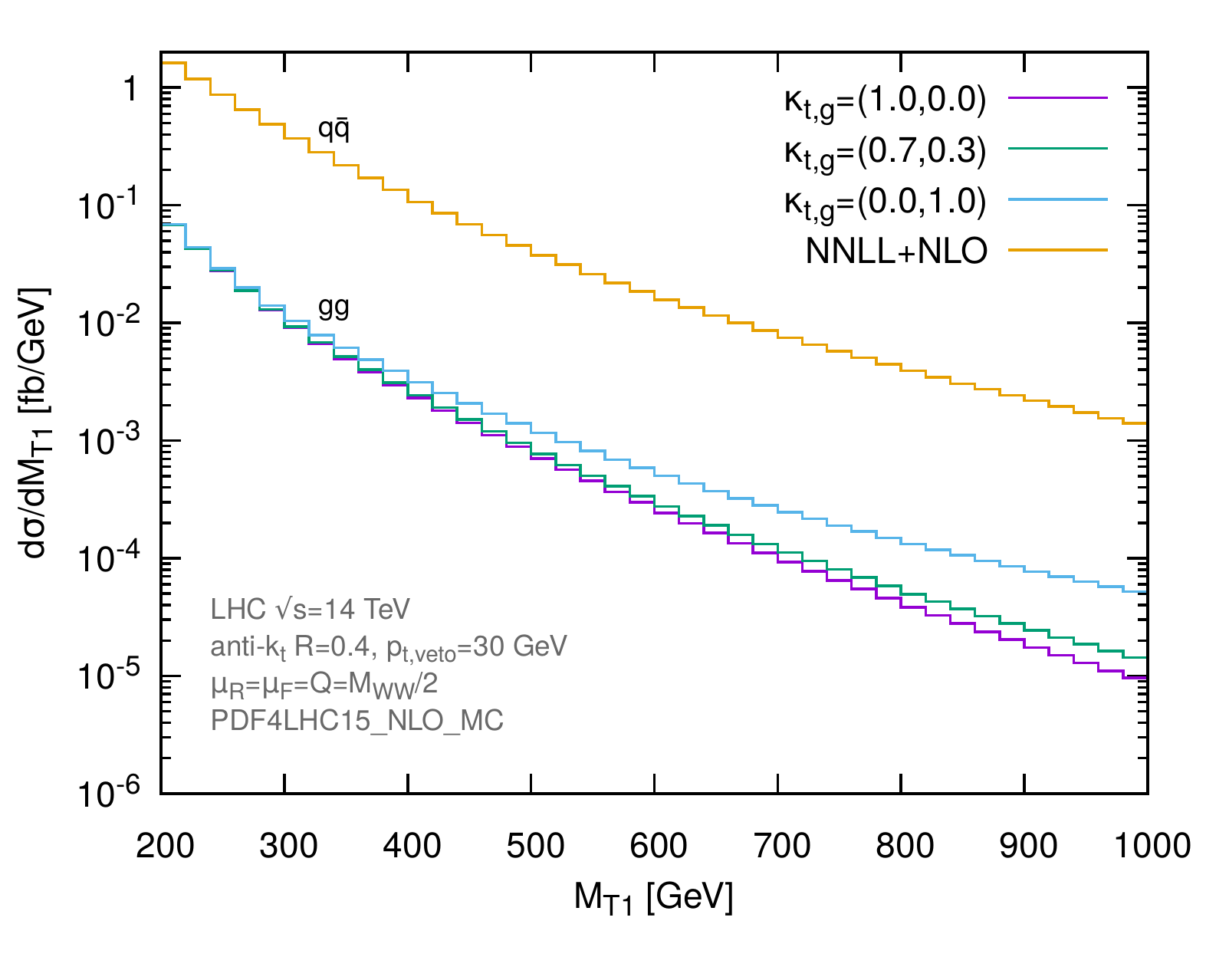}      
    \end{center}
  \end{minipage}
  \begin{minipage}[r]{0.5\linewidth}
    \begin{center}
      \includegraphics[width=\textwidth]{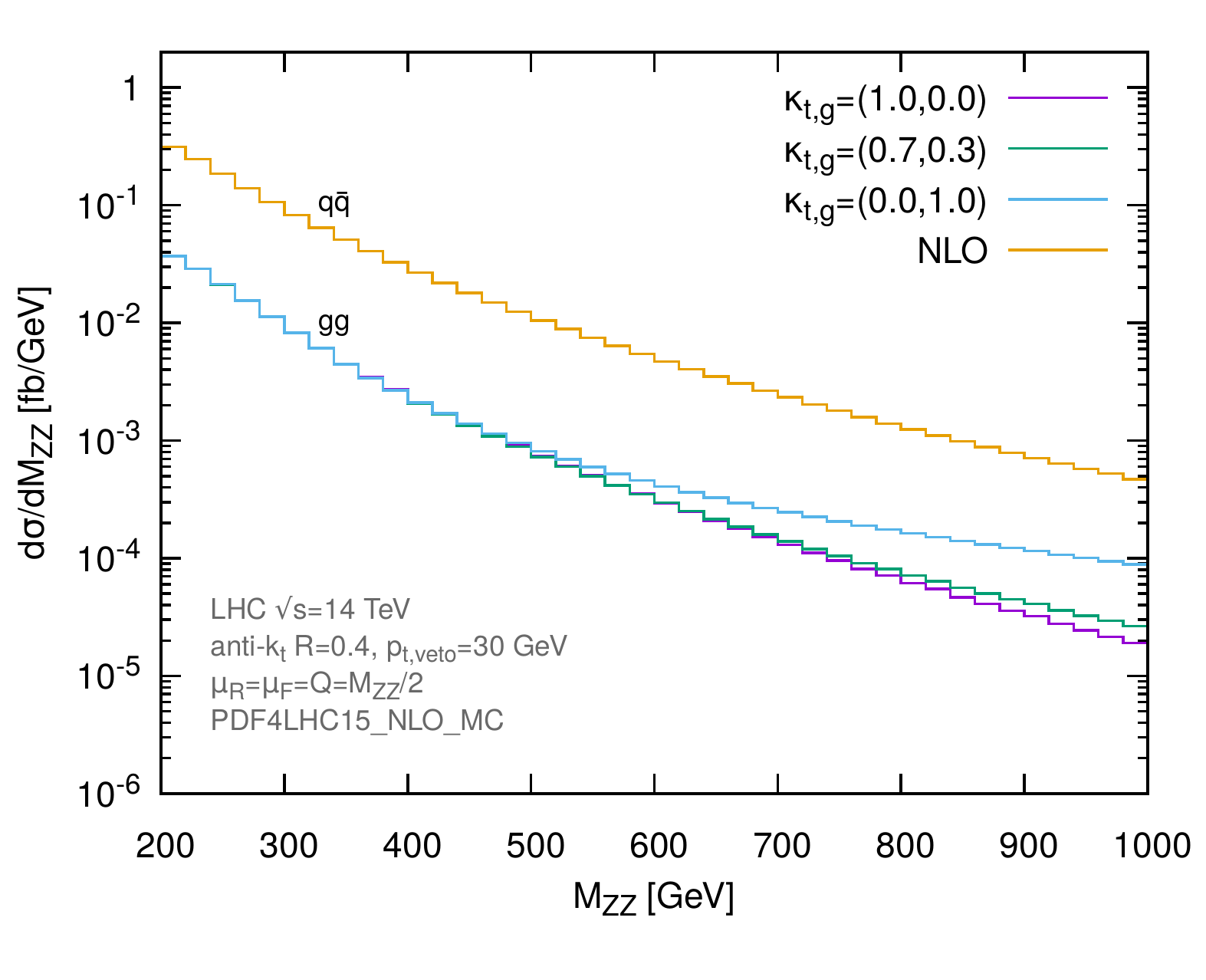}      
    \end{center}
  \end{minipage}
  \caption{Our best predictions for the differential distributions
    $d\sigma/dM_{T1}$ for $WW$ production with the experimental cuts
    in table~\ref{tab:cuts} (left) and $d\sigma/dM_{ZZ}$ for $ZZ$
    production with the cuts in ref.~\cite{Khachatryan:2015yvw}
    (right) for $q\bar q$ and $gg$ processes.}
  \label{fig:ww-vs-zz-14TeV}
\end{figure}

In this section, we compare the sensitivity of $WW$ and $ZZ$
production at HL-LHC ($\sqrt{s}=14\,$TeV, with $3\,$ab$^{-1}$ of
integrated luminosity) to the BSM operator considered in
eq.~\eqref{eq:BSM-operator}. Here we consider only the decay
$ZZ \to e^+ e^- \mu^+ \mu^-$. First we present the best predictions
that could be obtained with the theoretical tools considered here, for
a given choice of observables for the two processes. For $WW$ we
choose $M_{T1}$ in eq.~(\ref{eq:MT1}), and our best prediction is NNLL
for $q\bar q\to WW$ and NLL for $gg\to WW$. For $ZZ$ we consider
$M_{ZZ}$, and our best prediction is NLO for $q\bar q\to ZZ$ and LO
for $gg\to ZZ$. Note that the accuracy of the predictions for
$q\bar q$ annihilation for both $WW$ and $ZZ$ production can be
improved to include the most recent NNLO calculations of
refs.~\cite{Gehrmann:2014fva,Cascioli:2014yka}.  For gluon fusion,
full NLO corrections have yet to be calculated, although approximate
results are
available~\cite{Caola:2015ila,vonManteuffel:2015msa,Caola:2015psa,Caola:2015rqy,Caola:2016trd,Campbell:2016ivq,Grazzini:2018owa}. While
the inclusion of NNLO corrections to $ZZ$ is straightforward, and can
be obtained by running the code {\tt
  MATRIX}~\cite{Grazzini:2015hta,Kallweit:2018nyv,Grazzini:2016ctr,Grazzini:2017mhc,Grazzini:2016swo,Grazzini:2017ckn},
the use of NNLO corrections to $WW$ requires matching of fixed-order
predictions to the NNLL resummation. Although this can be achieved by
interfacing the NNLL resummation to {\tt MATRIX}, it is technically
more involved than the simple procedure described in
section~\ref{sec:background}. Therefore, we leave matching to NNLO to
future work. The differential distributions in $M_{T1}$ and $M_{ZZ}$
are shown in figure~\ref{fig:ww-vs-zz-14TeV}.
We observe that, in the $q\bar q$ channel, the cross section
$d\sigma/dM_{T1}$ with a jet-veto is comparable to the cross section
$d\sigma/dM_{ZZ}$ where no jet veto is applied. We note that, even with
a jet veto, the $q\bar q$ background is much larger in the $WW$
case. Therefore, we naively expect $WW$ to perform slightly worse than
$ZZ$ for exclusion of BSM effects.

To be more quantitative, we generate exclusion plots for a range of
values of the parameters $\kappa_t$ and $\kappa_g$ entering the
Lagrangian of eq.~(\ref{eq:BSM-model}). To do this we ask ourselves
how likely it is that predictions corresponding to different values of
$(\kappa_t, \kappa_g)$ are compatible with data that agree with the
SM.  Quantitatively, given a value of $(\kappa_t,\kappa_g)$, we
compute $n_i(\kappa_t,\kappa_g)$, the expected number of events in bin
$i$ of the distribution in a suitable leptonic
observable. Specifically, we choose $M_{T1}$ for $WW$ production and
$M_{ZZ}$ for $ZZ$.  Given a set of data points
$\{n_i\}_{i=1,\dots,N}$, and a given value of $(\kappa_t,\kappa_g)$,
we define
\begin{equation}
  \label{eq:chi2}
  \chi^2(\kappa_t,\kappa_g)\equiv \sum_i \frac{(n_i(\kappa_t,\kappa_g) - n_i)^2}{n_i}\,,
\end{equation}
and from that we construct our test statistic
\begin{equation}
  \label{eq:dchi2}
  \Delta\chi^2(\kappa_t,\kappa_g)\equiv \chi^2(\kappa_t,\kappa_g)-\chi^2(\hat\kappa_t,\hat\kappa_g)\,,
\end{equation}
where $(\hat\kappa_t,\hat\kappa_g)$ are the values of
$(\kappa_t,\kappa_g)$ that minimise $\chi^2(\kappa_t,\kappa_g)$.  This
test statistic is a good approximation to the usual log-likelihood
ratio for counting experiments~\cite{Tanabashi:2018oca} in the limit
of a large number of events, and in the assumption that there are no
correlations between bins. Assuming $n_i(\kappa_t,\kappa_g)$ is the
expected number of events, in the denominator of eq.~\eqref{eq:chi2}
we can approximate $n_i\simeq n_i(\kappa_t,\kappa_g)$. Therefore,
$\Delta\chi^2(\kappa_t,\kappa_g)$ is asymptotically distributed
according to a chi-squared distribution with two degrees of freedom
(see e.g.~\cite{Hocker:2001xe}), which we denote by
$f(\Delta \chi^2(\kappa_t,\kappa_g)\mid\kappa_t,\kappa_g)$.

We now consider data $\{n_i\}_{i=1,\dots,N}$ generated in such a way
that the expected number of events in each bin is the ``central'' SM
prediction, corresponding to $\mu_R=\mu_F=Q=M_{WW}/2$ for $WW$ and
$\mu_R=\mu_F=M_{ZZ}/2$ for $ZZ$, which we denote with $\bar
n_i(1,0)$. This constitutes our ``background-only'' hypothesis. We now
set exclusion limits in the $(\kappa_t,\kappa_g)$ plane using the
median significance~\cite{Tanabashi:2018oca,Cowan:2010js}, assuming
those data, with which one reject the hypothesis corresponding to each
value of $(\kappa_t,\kappa_g)$ (our ``signal'' hypothesis). More
precisely, for each value of $(\kappa_t,\kappa_g)$, we construct the
distribution in $\Delta\chi^2(\kappa_t,\kappa_g)$ under the assumption
of the background-only hypothesis, which we denote by
$f(\Delta \chi^2(\kappa_t,\kappa_g)\mid 1,0)$. We then compute the
median of that distribution, which we denote with
$\Delta\chi^2_{\rm med}(\kappa_t,\kappa_g)$. The $p$-value for each
$(\kappa_t,\kappa_g)$ is given by
\begin{equation}
  \label{eq:p-value}
  p(\kappa_t,\kappa_g)=\int^\infty_{\Delta\chi^2_{\rm med}(\kappa_t,\kappa_g)}
\!\!\!\!\!\!\!\!\!\!\!\!\!\!\!\!
  f(\Delta \chi^2\mid \kappa_t,\kappa_g)\,d(\Delta \chi^2) \,,
\end{equation}
and we exclude at the 95\% confidence level all $(\kappa_t,\kappa_g)$
such that $p(\kappa_t,\kappa_g)<0.05$. In practice, we have binned the
variables $M_{T1}$ and $M_{ZZ}$ in such a way that, when computing
$\Delta\chi^2_{\rm med}(\kappa_t,\kappa_g)$, in the denominator of
eq.~\eqref{eq:chi2} we can always approximate $n_i$ with
$\bar n_i^{q\bar q}$, the number of events obtained using central
scales and the $q\bar q$ subprocess only.

\begin{figure}[htbp]
  \centering
      \includegraphics[width=\textwidth]
      {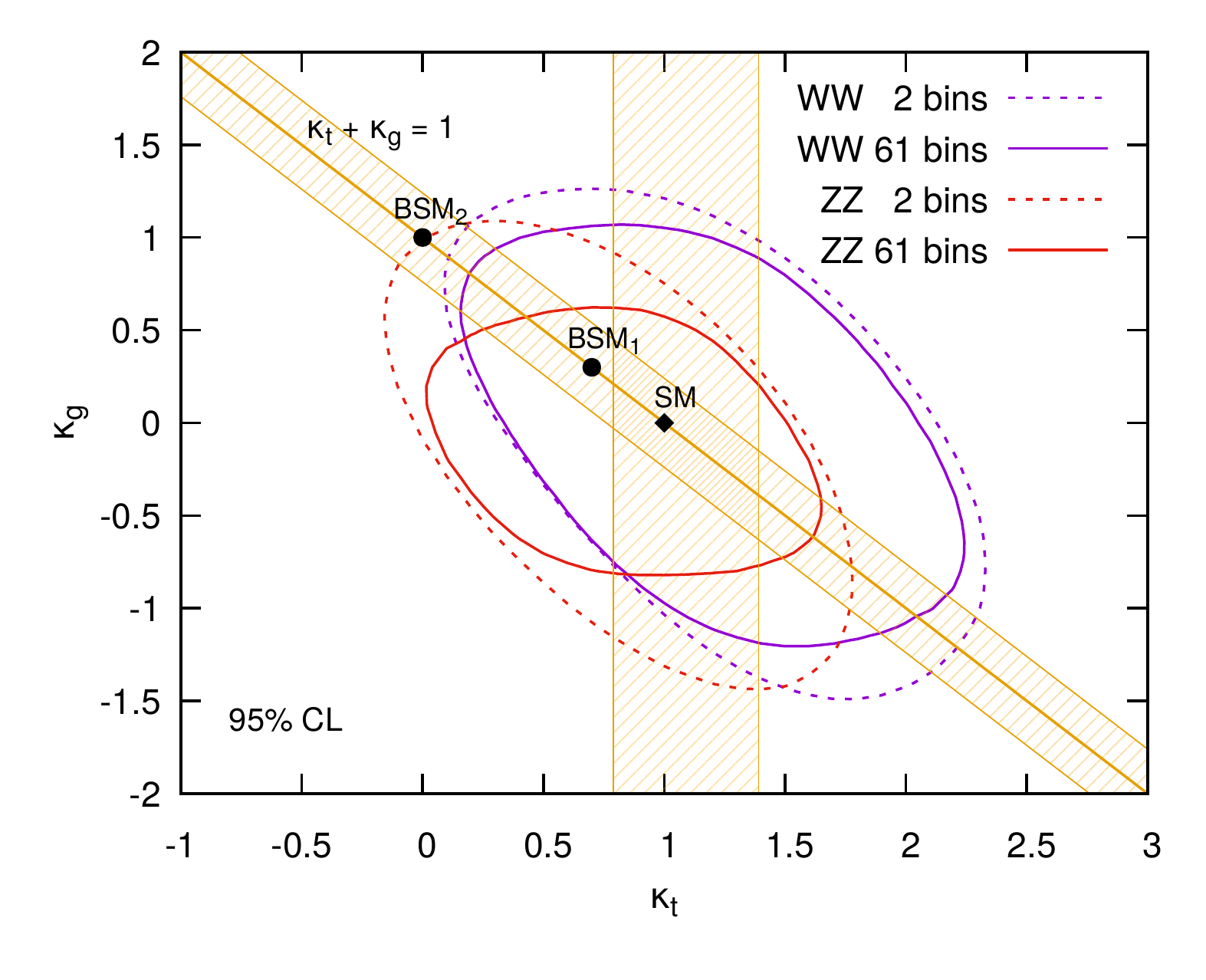}
      \vspace{-.5cm}
    \caption{Exclusion contours at 95\% level for $WW$ and $ZZ$
      production. See the main text for details.}
  \label{fig:MT1-MZZ-cen-contour}
\end{figure}

We first consider the case in which the expected number of events for
the signal hypothesis corresponds to $\bar n_i(\kappa_t,\kappa_g)$. We
have examined two cases, both corresponding to di-boson invariant
masses above the Higgs mass, so as to ensure to have complementary
information with respect to Higgs cross sections. In one case, we have
considered only two bins, a low-mass bin
($200\,\mathrm{GeV}<M_{T1},M_{ZZ}<400\,$GeV) and a high-mass bin containing the
rest of the distributions. The low-mass bin is more sensitive to
$\kappa_t$, and the high-mass bin to $\kappa_g$. The corresponding
exclusion regions in the $(\kappa_t,\kappa_g)$ plane are bounded by the
dashed contours in Fig.~\ref{fig:MT1-MZZ-cen-contour}.
We see that $WW$ is complementary to $ZZ$ for low values of
$\kappa_t$, whereas the sensitivity to $\kappa_g$ of $ZZ$ is
larger. This can be understood from
figure~\ref{fig:ww-vs-zz-14TeV}. Note that, despite the fact that the
$WW$ cross-section is larger, the presence of the jet veto kills a
good fraction of the $gg$ signal, with the net effect that its
cross-section decreases with increasing $M_{T1}$. In the $ZZ$ case,
where there is no suppression due to a jet veto, the contact
interaction driven by $\kappa_g$ is fully effective, and makes the
$gg$ signal flatter with respect to the $q\bar q$ background, thus
giving a larger sensitivity to $\kappa_g$. We gain sensitivity by
considering a greater number of bins. For instance, we have considered
60 bins equally spaced from $200\,$GeV to $1400\,$GeV, and an extra
bin containing the distribution with larger values of $M_{T1}$ or
$M_{ZZ}$. The corresponding exclusion contours are the solid lines in
Fig.~\ref{fig:MT1-MZZ-cen-contour}.  For reference, we also plot the
line $\kappa_t+\kappa_g=1$, and three points corresponding to the SM,
and the scenarios BSM$_{1}$ and BSM$_2$ considered in the previous
section. We also draw bands corresponding to 95\% confidence-level
bounds on $\kappa_t+\kappa_g$ and $\kappa_t$ obtained from
ref.~\cite{ATLAS:2018doi}. These give more stringent constraints than
our observables, which have nevertheless complementary sensitivity,
since the analysis of ref.~\cite{ATLAS:2018doi} probes regions of
di-boson invariant masses that we do not consider here. Also, having
full control of theoretical predictions for both the signal and the
background, our procedure is suitable for optimisation of both the
observables and the binning procedure, and is open to improvements
of the theoretical predictions. 

The exclusion contours we have obtained so far do not take into
account theoretical uncertainties.
Including theoretical uncertainties, the true theory value
$n_i(\kappa_t,\kappa_g)$ will differ from its central prediction
$\bar n_i(\kappa_t,\kappa_g)$ by some theoretical error $\delta_i$,
taken to lie in some interval $\Delta_i$.
In every bin, $n_i(\kappa_t,\kappa_g)$ will be the sum of a
contribution $n_i^{(q\bar q)}$ arising from quark-antiquark
annihilation, and a contribution $n_i^{(gg)}(\kappa_t,\kappa_g)$
arising from gluon fusion. Denoting by
$\Delta^{(gg)}_i(\kappa_t,\kappa_g)$ and $\Delta_i^{(q\bar q)}$ the
theoretical uncertainties on (respectively)
$n_i^{(gg)}(\kappa_t,\kappa_g)$ and $n_i^{(q\bar q)}$, and considering
the fact that these predictions correspond to completely uncorrelated
processes, we take the theoretical uncertainty on
$n_i(\kappa_t,\kappa_g)$ to be given by
\begin{equation}
  \label{eq:Deltai}
  \Delta_i(\kappa_t,\kappa_g)= \sqrt{\left(\Delta^{(gg)}_i(\kappa_t,\kappa_g)\right)^2+\left(\Delta_i^{(q\bar q)}\right)^2}\,.
\end{equation}
Therefore, the $\chi^2$ corresponding to a given value of $(\kappa_t,\kappa_g,\vec \delta\equiv \{\delta_1,\delta_2,\dots\})$ is given by
\begin{equation}
  \label{eq:chi2-exp}
  \chi^2_{\rm exp}(\kappa_t,\kappa_g,\vec \delta)\equiv \sum_i \frac{(\bar n_i(\kappa_t,\kappa_g)+\delta_i - n_i)^2}{n_i}\,. 
\end{equation}
In order to estimate the impact of theoretical uncertainties on our
sensitivity contours, we adopt the approach of
ref.~\cite{Hocker:2001xe}, and add to $\chi^2_{\rm exp}$ a Gaussian
``theory term'',
with a width $\Delta_i(\kappa_t,\kappa_g)/2$, as follows:
\begin{equation}
  \label{eq:chi2-th}
  \chi^2_{\rm th}(\kappa_t,\kappa_g,\vec \delta)\equiv\sum_i \frac{\delta_i^2}{\left(\Delta_i(\kappa_t,\kappa_g)/2\right)^2}\,.
\end{equation}
The test statistic corresponding to $(\kappa_t,\kappa_g)$ is then
obtained by profiling with respect to $\vec \delta$, i.e.\ computing
\begin{equation}
  \label{eq:chi2-final}
  \chi^2(\kappa_t,\kappa_g)\equiv \min_{\vec\delta} \left[\chi^2_{\rm exp}(\kappa_t,\kappa_g,\vec \delta)+\chi^2_{\rm th}(\kappa_t,\kappa_g,\vec \delta)
\right].
\end{equation}
For $\chi^2_{\rm exp}$ and $\chi^2_{\rm th}$ as in (\ref{eq:chi2-exp})
and (\ref{eq:chi2-th}) this gives
\begin{equation}
  \label{eq:chi2-final1}
  \chi^2(\kappa_t,\kappa_g) = \sum_i \frac{ (\bar n_i(\kappa_t, \kappa_g) - n_i)^2}{n_i + \left(\Delta_i(\kappa_t, \kappa_g)/2\right)^2 } .
\end{equation}
In other words, for a Gaussian theory term our treatment is equivalent
to the common prescription to combine theoretical and experimental
errors in quadrature.\footnote{In fact, (\ref{eq:chi2-th}) itself can
  similarly be obtained as follows: (i) introduce separate
  $\delta_i^{(gg)}$ and $\delta_i^{(q \bar q)}$ parameters for the two
  components of $n_i(\kappa_t, \kappa_g)$, (ii) add separate Gaussian
  theory terms for the former, of respective widths
  $\Delta_i^{(gg)}/2$ and $\Delta_i^{(q \bar q)}/2$, (iii) define
  $\delta_i = \delta_i^{(gg)} + \delta_i^{(q \bar q)}$ and rewrite the
  $\chi^2$ in terms of $\delta$ and $\delta_i^{(q \bar q)}$, (iv)
  profile (minimise) the $\chi^2$ with respect to
  $\delta_i^{(q \bar q)}$. This again gives the expression
  (\ref{eq:chi2-final1}).} With our choice of bins, we can approximate
$\Delta_i(\kappa_t, \kappa_g)\simeq \Delta_i^{(q\bar q)}$.

\begin{figure}[!htbp]
  \begin{center}
    \includegraphics[width=\textwidth]{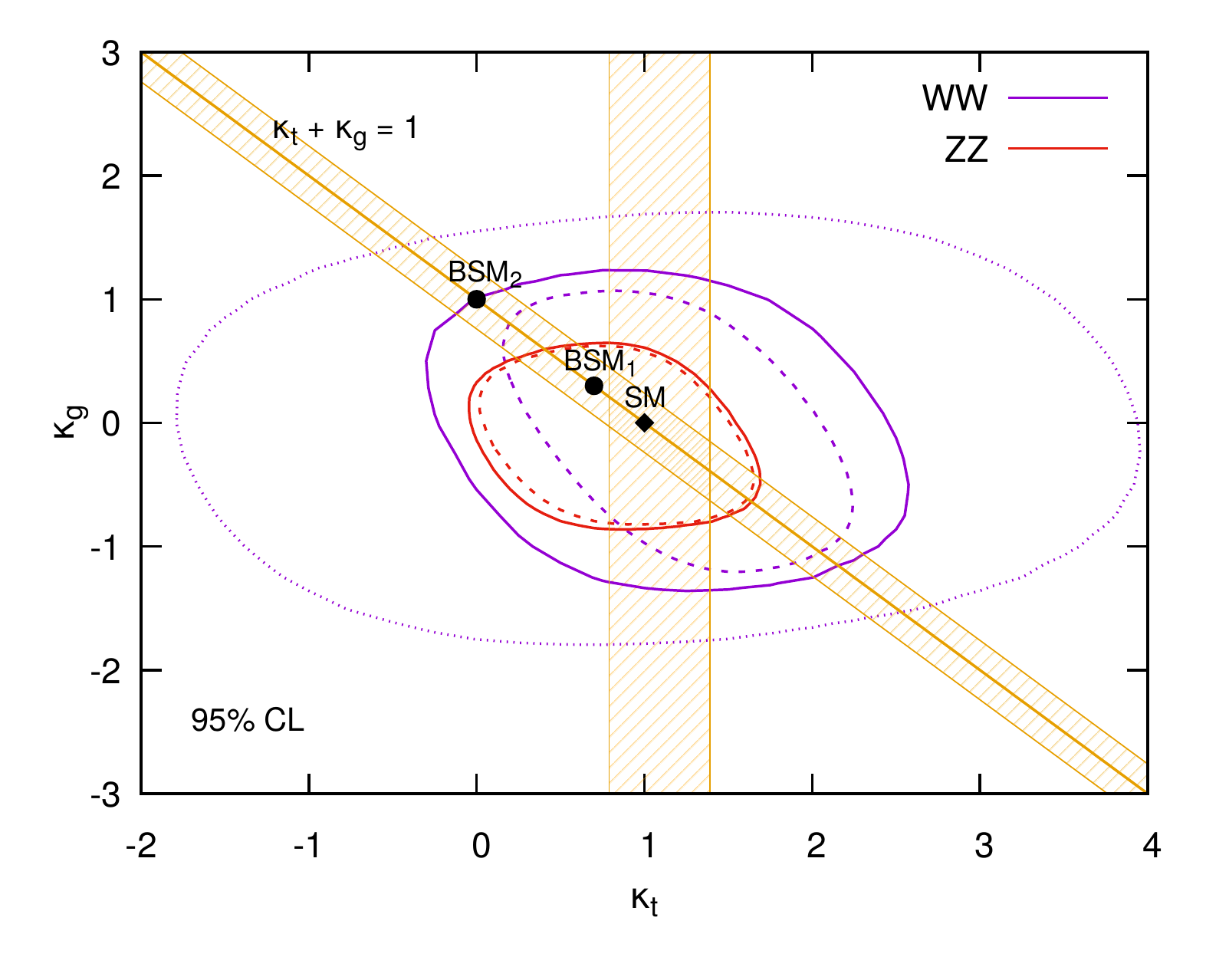}
  \end{center}
  \vspace{-1.2cm}
  \caption{Exclusion contours at 95\% level for $WW$ and $ZZ$
    production, corresponding to different ways of estimating
    theoretical uncertainties, see the main text for details.}
  \label{fig:MT1-MZZ-theory}
\end{figure}

Before presenting sensitivity contours including theory uncertainties,
it is worth comparing the impact of statistical and theoretical
uncertainties. In the case of $WW$ production, theory uncertainties
differ according to whether we use the efficiency method described in
appendix~\ref{sec:th-uncerts}, or we just perform 9-point scale
variations in the resummed cross section. In the former case, as can
be seen from Fig.~\ref{fig:SM-MC}, relative theory uncertainties are
of order 40\%, whereas in the latter they are of order 10\%, with a
mild dependence on $M_{WW}$. In both cases then $\Delta_i^{(q\bar q)}$
roughly scales like $n_i$. Therefore, by looking at the denominator of
eq.~\eqref{eq:chi2-final1}, we see that in the bins with larger $n_i$,
theory uncertainties will dominate over statistical uncertainties
($\sim \sqrt n_i$), and hence these bins have very little power to
constrain $(\kappa_t, \kappa_g)$. In the case of $ZZ$, theory
uncertainties are smaller, around 5\%, so all bins retain their
constraining power. This is illustrated in
Fig.~\ref{fig:MT1-MZZ-theory}.
All contours have been obtained with 61 bins, as explained above. The
outer contour (dotted) corresponds to $WW$ production with theory
uncertainties estimated with the JVE method. As explained in
sec.~\ref{sec:results}, the method is probably overly conservative,
and the corresponding contour cannot compete with the constraints from
$ZZ$ production. Note in particular that large theoretical
uncertainties affect mostly the bins with lowest values of $M_{T1}$,
which are the most sensitive to $\kappa_t$. This explains why the JVE
contour is so wide compared to the others. The solid contours
correspond to uncertainties obtained with the appropriate scale
variations, both for $WW$ and for $ZZ$. Based on previous works on
Higgs production with a
jet-veto~\cite{Banfi:2015pju,Banfi:2012jm,Banfi:2012yh}, we believe
that scale variations for $WW$ give a realistic estimate of the best
theoretical uncertainties that could be obtained with a matching to
NNLO with the JVE method. We see that, taking into account theory
uncertainties at the currently achievable accuracy, $WW$ does not have
complementary constraining power with respect to $ZZ$. However, the
dashed curves, corresponding to all predictions fixed at their central
value without theory uncertainties, show that $WW$ might compete with
$ZZ$. We have therefore determined the necessary accuracy on $WW$
production such that one obtains a comparable sensitivity with
$ZZ$. First, we have observed that, in the case of $ZZ$, adding the
NNLO contribution to $q\bar q$ does not improve the overall theory
accuracy, due to missing higher orders in the $gg$ channel. So we
assume that the uncertainties on $ZZ$ production will remain the NLO
ones, i.e. around 5\%.
\begin{figure}[htbp]
  \begin{center}
    \includegraphics[width=\textwidth]{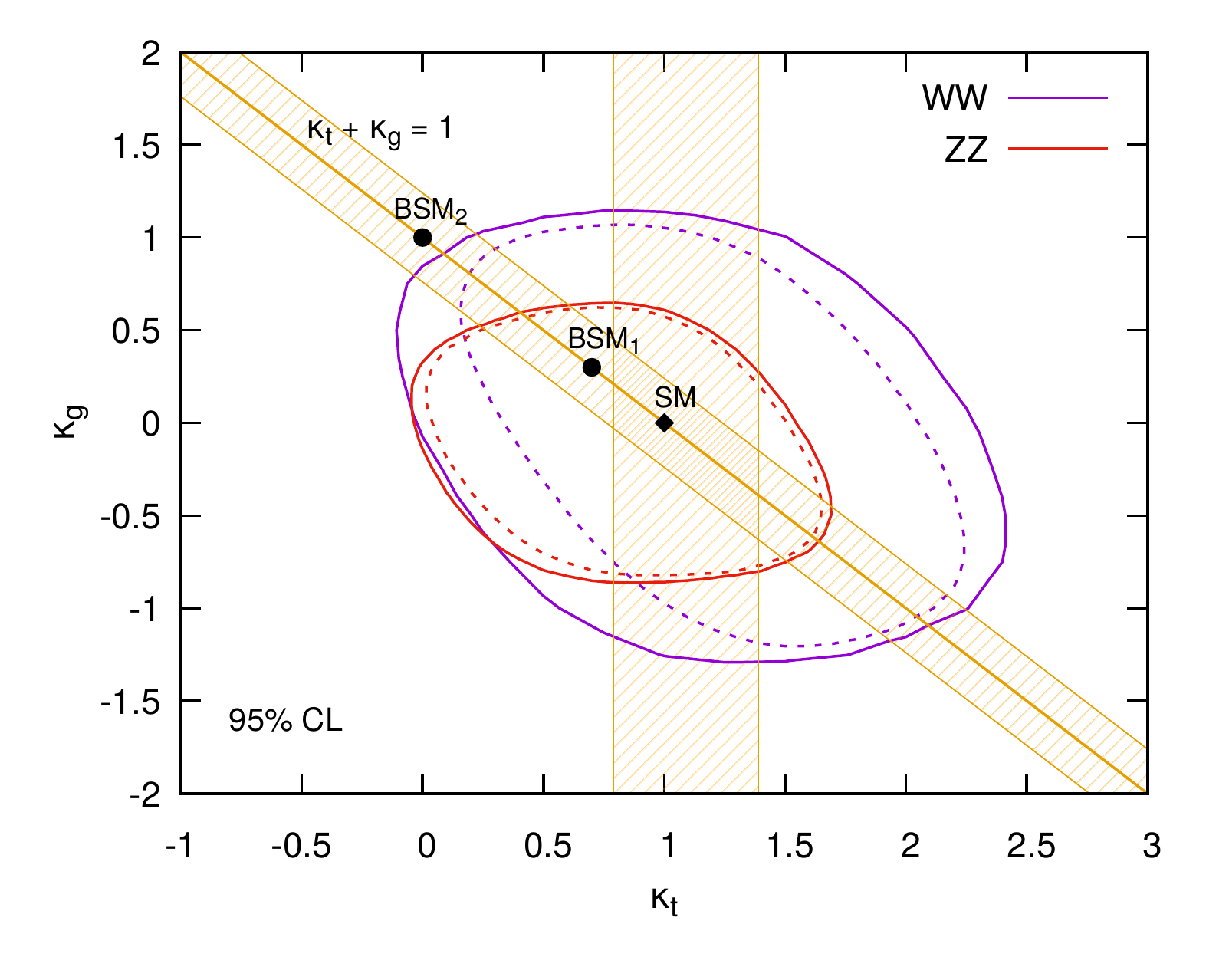}      
  \end{center}
  
  \vspace{-1.2cm}
  
  \caption{Exclusion contours at 95\% level for $WW$ and $ZZ$
    production, corresponding to an optimistic reduction of
    theoretical uncertainties, see the main text for details.}
  \label{fig:MT1-MZZ-projection}
\end{figure}
The solid contour for $WW$ in Fig.~\ref{fig:MT1-MZZ-projection}
corresponds to an estimated theoretical uncertainty of 3\% in every
bin, which is approximately the one you need for $WW$ to be
competitive with current $ZZ$ predictions. Based again on previous
work on Higgs production~\cite{Banfi:2015pju}, such an uncertainty
could be reached by matching NNLL resummation to a future NNLO
calculation for $WW$ plus one jet, and maybe even further decreased
after an N$^3$LL resummation. We note that improving $ZZ$ predictions
hardly offers any stronger constraint. However, improved predictions
for the $gg$ channel, both for $WW$ and $ZZ$, might move the central
prediction, and may open up further space for constraints.

We conclude this section with a comment on the actual implementation
of the calculation of $\chi^2(\kappa_t,\kappa_g)$. If we consider the
numerator of $\chi^2(\kappa_t,\kappa_g)$ in eqs.~\eqref{eq:chi2}
and~(\ref{eq:chi2-final1}), we see that it involves
$n_i^{(gg)}(\kappa_t,\kappa_g)$. This quantity is a second-order
polynomial in $\kappa_t$ and $\kappa_g$, arising from the square of
the matrix element
\begin{equation}
  \label{eq:Mgg}
  M^{(gg)}_\SM + M^{(gg)}_\BSM = \kappa_t M^{(gg)}_t + \kappa_g M^{(gg)}_g +  M^{(gg)}_c\,,
\end{equation}
where $M^{(gg)}_t$ and $M^{(gg)}_g$ are the contributions of the Higgs
produced via a top loop and a contact interaction respectively, and
$M^{(gg)}_c$ the remaining contributions, giving rise to the so-called
``continuum'' background.  The fact that we have full control over
$M^{(gg)}$ allows us to compute the coefficient of each power of
$\kappa_t$ and $\kappa_g$ separately, and once and for all. This is
crucial for an accurate calculation of
$\chi^2(\kappa_t,\kappa_g)$, because a naive implementation of
this quantity might involve cancellations between large numbers, whose
control requires Monte Carlo samples with large statistics.

\section{Conclusions}
\label{sec:conclusions}

We have studied the impact of a veto on additional jets on setting
limits on the coupling of a dimension-6 operator affecting $WW$
production. In the presence of such a veto, large logarithms of the
ratio of the maximum allowed jet transverse momentum $\ptjv$ and the
invariant mass of the $WW$ pair $M_{WW}$ have to be resummed at all
orders in QCD. These logarithmically enhanced contributions give rise
to the so-called Sudakov suppression of cross sections with respect to
naive Born-level predictions. The dimension-6 operator we considered
affects $WW$ production via gluon fusion, but does not affect $WW$
production via quark-antiquark annihilation, which stays unchanged
with respect to the SM. At Born level, the effect of this
operator amounts to a growth of the cross section at large values of
$M_{WW}$. Unfortunately, the suppression due to the jet-veto gets
larger with increasing $M_{WW}$. Also, such suppression affects gluon
fusion more than quark-antiquark annihilation due to the fact that
gluons radiate roughly twice as much as quarks, so vetoing radiation
off gluons cuts a larger portion of cross sections. Therefore,
enhancement due to a contact interaction and Sudakov suppression are
in competition.

To investigate quantitatively the impact of a jet veto on $WW$
production, we have devised a new method to interface resummed
predictions for the $gg$ and $q\bar q$ channels to
fixed-order QCD event generators. This procedure provides events that
are fully differential in the decay products of the $WW$ pair, so that
suitable acceptance cuts can be applied. The method involves minimal
modifications of the ingredients already present in fixed-order event
generators, and can be applied to the production of any colour
singlet. In particular, we have implemented the procedure in the
fixed-order program \mcfm, which resulted in the code we called
\mcfmre, a Resummation Edition of \mcfm.

Our program \mcfmre\ has been used to produce differential cross
sections for $WW$ production with a simplified version of the ATLAS
acceptance cuts, both in the SM, and including BSM effects induced by
the aforementioned dimension-6 contact interaction. The main message
is that, with the value of $\ptjv$ used in current analyses, Sudakov
suppression effects dominate over the enhancement produced by a contact
interaction, so that deviations from the Standard Model are in general
quite small for reasonable values of the strength of the contact
interaction.

We have compared our results with those obtained from a number of
parton-shower event generators, and we have found very good agreement.
We have used parton-shower event generators to estimate effects that
cannot be not be taken into account by our analytical calculation, and
found that they have a small impact, well within our theory
uncertainties.  We emphasise that our predictions have the
computational cost of a Born-level event generator, and provide full
analytical control of theoretical uncertainties. Our predictions are
also in agreement, within uncertainties, with those obtained by
interfacing a SCET calculation with the same formal accuracy with
\amcatnlo.

We produced projections for the sensitivity to the considered BSM
effects for HL-LHC, and compared with what could be obtained using
$ZZ$ production, which is not affected by the presence of a jet
veto. We have found that $WW$ has complementary sensitivity, provided
it is possible to reduce theory uncertainties below 3\%. This could be
achieved by both matching current resummed predictions with a future
calculation of $WW$ plus one jet at NNLO, and improving the
resummation to achieve N$^3$LL accuracy. We hope this work encourages
further theoretical work in both directions. We remark that the main
advantage of using \mcfmre\ for such studies compared to parton-shower
event generators is that we have access to amplitudes, so we can
compute separately all terms contributing to square matrix elements,
in particular interference terms which can be computed separately with
an arbitrary numerical accuracy.

We have found that, with the current acceptance cuts, the observables
we have considered are not yet competitive with Higgs total cross
sections, although they do provide additional information. However,
our code does provide an accurate and fast tool to explore different
choices of cuts and observables, so could be used for further studies
in this direction. In fact, with minimal modifications, it can also
produce predictions with event-by-event jet vetoes, as proposed
in~\cite{Pascoli:2018rsg,Pascoli:2018heg}. Furthermore, our code is
open to the implementation of other models of new physics affecting
the production of a colour singlet.

Last, our code is the only implementation of the jet-veto resummation
of ref.~\cite{Banfi:2015pju} that is fully exclusive in the decay
products of a colour singlet, so it can be used for precision
determinations of Standard Model parameters, notably those
characterising the Higgs boson.

\acknowledgments We are grateful to Tilman Plehn for suggesting a
reference and to Christian Reuschle for extensive advice on
configuring \herwig{} as well as Keith Hamilton, Simon Pl\"atzer and
Peter Richardson for useful comments on \herwig{}-related issues. We
also acknowledge useful discussions on statistics with Luiz Vale
Silva, and on the effect of rapidity cuts on jet-veto resummations with
Johannes Michel and Frank Tackmann.  N.K.\ would like to thank CERN for
hospitality and partial financial support through the CERN Theory
Institute on \emph{LHC and the Standard Model: Physics and Tools}.
This work was supported in part by STFC grant ST/P000738/1.

\appendix

\section{Collection of relevant formulae}
\label{sec:formulae}

In this appendix we report the explicit expressions that we have
implemented in \mcfm\ to achieve NLL and NNLL resummation of the cross
section for the production of a colour singlet with a jet-veto. This
discussion is of a technical nature, and we assume that the reader is
familiar with the details of the jet-veto resummations performed
in refs.~\cite{Banfi:2012yh,Banfi:2012jm,Banfi:2015pju}.

In general, we consider the production of a colour singlet of
invariant mass $M$, for instance a Higgs, a $Z$ boson, or a pair of
$W$ bosons. At Born-level, this proceeds via either $q \bar q$
annihilation or gluon fusion. We then compute the cross section
$d\sigma_{\proc}/(dM^2 d\Phi_n)$, with $\proc=q\bar q,gg$, fully
differential in the phase space of the decay products of the colour
singlet. Given their momenta $q_1,q_2,\dots,q_n$, and incoming
momenta $p_1$ and $p_2$, the phase space $d\Phi_n$ is defined as
\begin{equation}
  \label{eq:dPhin}
  d\Phi_n = \prod_{i=1}^n \frac{d^3 \vec q_i}{(2\pi)^3 2 E_i}(2\pi)^4\delta\left(p_1+p_2-\sum_{i=1}^n q_i\right)\,,
\end{equation}
with $E_i$ and $\vec q_i$ the
energy and three-momentum of particle $q_i$.

Any prediction for $d\sigma_{\proc}/(dM^2 d\Phi_n)$ depends on the
renormalisation scale $\mu_R$ at which we evaluate the strong coupling
$\alpha_s$, as well as the factorisation scale $\mu_F$ at which we
evaluate the PDFs. Both scales are typically set at values of order
$M$. Furthermore, in the presence of a jet-veto,
$d\sigma_{(\proc)}/(dM^2 d\Phi_n)$ is affected by large logarithms
$L\equiv\ln(M/\ptjv)$, with $\ptjv$ the maximum allowed transverse
momentum of the observed jets. When resumming such logarithms at all
orders, our predictions become functions of $\tilde L$, defined as
\begin{equation}
  \label{eq:Ltilde}
  \tilde L = \frac{1}{p}\ln\left(\left(\frac{Q}{\ptjv}\right)^p+1\right)\,.
\end{equation}
The quantity $\tilde L$ is such that for large $\ptjv$,
$ \tilde L\to 0$, which implements the fact that, in this regime,
there are no large logarithms to be resummed. Also, at small $\ptjv$,
$\tilde L \simeq \ln(Q/\ptjv)$, so in fact we resum logarithms of the
ratio of $\ptjv$ and the so-called resummation scale $Q$. The three
scales $\mu_R,\mu_F,Q$ are handles that we will use to estimate
theoretical uncertainties, as explained in
app.~\ref{sec:th-uncerts}. The power $p$ determines how fast the
resummation switches off at large $\ptjv$. We choose $p=5$, as in
refs.~\cite{Banfi:2015pju,Banfi:2012jm,Banfi:2012yh}.

\subsection{NLL resummation}
\label{sec:NLL}

At NLL accuracy, the distribution $d\sigma_\proc/(dM^2 d\Phi_n)$ is given by
\begin{equation}
  \label{eq:dsigma-NLL}
  \frac{d\sigma^{\rm NLL}_\proc}{dM^2 d\Phi_n} = {\mathcal
    L}_{\proc}^{(0)}(\tilde L,M) \,e^{ \tilde L g_1(\as  \tilde L)+g_2(\as  \tilde L)}\,,\qquad \alpha_s=\alpha_s(\mu_R)\,.
\end{equation}
Explicit expressions for the functions $g_1,g_2$ can be found in the
supplemental material of ref.~\cite{Banfi:2012jm}. The NLL ``luminosity'' ${\mathcal L}_{\proc}^{(0)}(L,M)$ is given by
\begin{equation}
  \label{eq:L0}
  \mathcal{L}^{(0)}_{\proc}\left(L, M\right) \equiv \sum_{i,j}\int dx_1 dx_2 \, \left| M^{(\proc)}_{ij}\right|^2 \,\delta(x_1 x_2 s - M^2)
  f_i \left( x_1, \mu_F e^{-L} \right) f_j \left( x_2, \mu_F e^{-L} \right)\,.
\end{equation}
In the above expression, $M^{(\proc)}_{ij}$ is the Born-level
amplitude for the production of the colour singlet via annihilation of
the two partons $i$ and $j$, and $f_{i,j}$ is the density of parton $i,j$ in
the proton.

Given any Born-level event generator, the recipe to implement the NLL
resummation of eq.~\eqref{eq:dsigma-NLL} is straightforward:
\begin{enumerate}
\item change the factorisation scale $\mu_F$ provided by the generator
  to  $\mu_F e^{-\tilde L}$;
\item multiply the weight of every event by a factor
  $\exp\left[\tilde L g_1(\alpha_s \tilde L)+g_2(\alpha_s \tilde
    L)\right]$.
\end{enumerate}
Note that, if $\ptjv$ is fixed, and we do not integrate over different
values of $M^2$, both operations can be performed without touching the
Born-level generator code. In fact, many programs allow a change in
the factorisation scale by a constant factor. Also, the rescaling of
the weight can be performed by the analysis routines that produce
histograms for physical distributions. In our implementation, since we
do want to integrate over $M^2$, we have implemented the change in
factorisation scale inside the \mcfm\ code.

Another advantage we have in using \mcfm\ is that it gives us access
to the matrix elements in a form that is human readable. This is
particularly useful in case one wishes to separate contributions from
different parts of the matrix element, for instance a possible BSM contribution
from that of the SM background. We consider here the case of $WW$
production via gluon fusion, but the argument applies to other
processes as well. There, the Born-level matrix element has the
form $M^{(gg)}=M_\SM^{(gg)}+M_\BSM^{(gg)}$, where $M_\SM^{(gg)}$ is
the SM amplitude, and $M_\BSM^{(gg)}$ a BSM contribution. For each
phase space point, we can then isolate individual contributions to the
luminosity by computing separately each term in the square
\begin{equation}
  \label{eq:Mgg-squared}
|M^{(gg)}|^2=|M_\SM^{(gg)}|^2+|M_\BSM^{(gg)}|^2+2\mathrm{Re}\left[M_\SM^{(gg)} \left(M_\BSM^{(gg)}\right)^*\right]\,.
\end{equation}
In the specific case, given the expression of $M^{(gg)}$ in
eq.~(\ref{eq:Mgg}), we compute the luminosity
$\mathcal{L}^{(0)}_{gg}\left(L, M\right)$ as follows
\begin{multline}
  \label{eq:Lgg-polynomial}
  \mathcal{L}^{(0)}_{gg}(L,M) = \kappa_t^2\, \mathcal{L}_{gg}^{(t^2)}(L,M) + \kappa_g^2
  \, \mathcal{L}_{gg}^{(g^2)}(L,M) + \kappa_t \kappa_g \, \mathcal{L}_{gg}^{(tg)}(L,M) + \\
  + \kappa_t \, \mathcal{L}_{gg}^{(tc)}(L,M) + \kappa_g \,
  \mathcal{L}_{gg}^{(gc)}(L,M) + \mathcal{L}_{gg}^{(c^2)}(L,M)\,,
\end{multline}
where we have used the notation
\begin{equation}
  \mathcal{L}_{gg}^{(i^2)}\left(L,M\right) = \int dx_1 dx_2 \, \left| M^{(gg)}_i \right|^2 \,\delta(x_1 x_2 s - M^2) 
  f_g \left( x_1, \mu_F e^{-L} \right) f_g \left( x_2, \mu_F e^{-L} \right)\,,
\end{equation}
with $i=t, g, c$, and
\begin{multline}
  \mathcal{L}_{gg}^{(ij)}\left(L,M\right) = \int dx_1 dx_2 \, 2 {\rm Re}\left[
    M^{(gg)}_i \left( M^{(gg)}_j \right)^* \right]\,\delta(x_1 x_2 s - M^2) \times \\
  \times f_g \left( x_1, \mu_F e^{-L} \right) f_g \left( x_2, \mu_F e^{-L}
  \right)\,,
\end{multline}
with $ij=tg,tc,gc$. Using these luminosities we can interpret
$\mathcal{L}_{gg}^{(0)}$ as a polynomial in the various $\kappa_i$,
and compute each coefficient separately. All one has to do then is to
reweight each phase-space point using the Sudakov exponent
$\exp\left[ \tilde L g_1(\as \tilde L) + g_2(\as \tilde L)
\right]$. In doing so, we have used the fact that the Sudakov exponent
depends only on the colour and kinematics of the incoming partons,
and therefore is the same for every single contribution to the
luminosity.

\subsection{NNLL resummation}
\label{sec:NNLL}

At NNLL accuracy, the cross section $d\sigma_{\proc}/(dM^2 d\Phi_n)$
with a jet veto is given by
\begin{multline}
  \label{eq:dsigma-NNLL}
  \frac{d\sigma^{\rm NNLL}_{\proc}(\pv)}{dM^2 d\Phi_n}= \left(
    {\mathcal L}_{\proc}^{(0)}(\tilde L,M)+{\mathcal
      L}_{\proc}^{(1)}(\tilde L,M)\right) \times \\ \times \left
    (1+\mathcal{F}_{\text{clust}}(R)+\mathcal{F}_{\text{correl}}(R)\right)
  \times e^{ \tilde L g_1(\as \tilde L)+g_2(\as \tilde L)+\frac{\as}{\pi} g_3(\as \tilde L)}\,,
\end{multline}
where the function $g_3$ can be found in ref.~\cite{Banfi:2012jm}.
The functions
$\mathcal{F}_{\text{clust}}(R),\mathcal{F}_{\text{correl}}(R)$ depend
on the jet radius $R$. Their expressions can be found in
ref.~\cite{Banfi:2012yh}. As for the NLL resummation,
$\alpha_s=\alpha_s(\mu_R)$. The remaining new ingredient for NNLL
resummation is the luminosity ${\mathcal L}_{\proc}^{(1)}(L,M)$,
defined as
\begin{align}
\label{eq:L1}
&   {\mathcal L}_{\proc}^{(1)}(L,M)  =  \sum_{i,j}\int
                           dx_1 dx_2 |M^{(\proc)}_{ij}|^2 \delta(x_1 x_2 s -
                M^2) 
 \frac{\alpha_s}{2\pi} \left[ {\cal H}^{(1)}_{\proc}f_i\!\left(x_1, \mu_F e^{-L}\right)
  f_j\!\left(x_2, \mu_F e^{-L}\right)  \notag \right. \\
  &+\left. \frac{1}{1-2 \alpha_s \beta_0  L}\sum_{k}\left(
  \int_{x_1}^1\frac{dz}{z} C_{ik}^{(1)}(z)
  f_k\!\left(\frac{x_1}{z}, \mu_F e^{-L}\right)
  f_j\!\left(x_2, \mu_F e^{-L}\right) +\{(x_1,i)\,\leftrightarrow\,(x_2,j)\}\right)\, \right]\,, 
\end{align}
with $\beta_0=(11 C_A-2 n_f)/(12\pi)$.  Using the conventions of
ref.~\cite{Banfi:2012jm}, we have
\begin{equation}
  \label{eq:H}
  \begin{split}
  \mathcal{H}_{q\bar q}^{(1)} & = H^{(1)} -2 C_F\left(\frac{3}{2}+\ln\frac{M^2}{Q^2}\right)\ln\frac{M^2}{Q^2}\,,\\
  \mathcal{H}_{gg}^{(1)} & = H^{(1)} -2 C_A\left(2\pi \beta_0 +\ln\frac{M^2}{Q^2}\right)\ln\frac{M^2}{Q^2}\,.
  \end{split}
\end{equation}
with $H^{(1)}$ the finite part of one-loop virtual corrections to the
process in question, e.g.\ $WW$ production through $q\bar q$
annihilation. The coefficients $C^{(1)}_{ij}$ depend on whether incoming
partons $i$ and $j$ are quarks/antiquarks ($q$) or gluons ($g$), and are given by:
\begin{equation}
  \label{eq:C}
  \begin{split}
  C^{(1)}_{qq}(z)& =C_F\left[(1-z)-\frac{\pi^2}{12} \delta(1-z)+\left(\frac{1+z^2}{1-z}\right)_+\ln\frac{Q^2}{\mu_F^2}\right]\,,\\
  C^{(1)}_{qg}(z)& = \frac{1}{2}\left[2  z (1-z)+(1-2z(1-z))\ln\frac{Q^2}{\mu_F^2}\right]\,,\\
  C^{(1)}_{gq}(z)& = C_F \left[z+\left(\frac{1+(1-z)^2}{z}\right)\ln\frac{Q^2}{\mu_F^2}\right]\,, \\
  C^{(1)}_{gg}(z)& = C_A\left[\left(2\pi\beta_0-\frac{\pi^2}{12}\right) \delta(1-z) +2\left(\frac{z}{(1-z)_+}+\frac{1-z}{z}+z(1-z)\right)\ln\frac{Q^2}{\mu_F^2}\right] \,. 
  \end{split}
\end{equation}
As explained in the previous section, the NLL luminosity
${\mathcal L}_{\proc}^{(0)}$ can be obtained from a Born-level event
generator. The function ${\mathcal L}^{(1)}_{\proc}$ represents a
correction to ${\mathcal L}_{\proc}^{(0)}$ of relative order
$\alpha_s$. Therefore, its implementation requires at least a NLO
generator. Any NLO event generator
includes the calculation of virtual corrections, as well as integrated
counterterms. This contribution, which we denote by
$d\sigma^{(1)}_{\proc, v+ct}/(d\Phi_n dM^2)$, has the same form as the
luminosity ${\mathcal L}^{(1)}_{\proc}$, but with PDFs evaluated at a
different factorisation scale, and different functions replacing
$\mathcal{H}^{(1)}_\proc$ and $C^{(1)}_{ij}(z)$. Its expression in
general depends on the way each process is implemented in the NLO
event generator. For instance, the implementation of $WW$ production
in the NLO program \mcfm\ follows from the general coding of the
production of a colour singlet, whose details can be found in
ref.~\cite{Campbell:2000bg}. Schematically,
\begin{align}
\label{eq:ds1-MCFM}
& \left( \frac{d\sigma^{(1)}_{\proc, v+ct}}{d\Phi_n dM^2}\right)_{\!\!\mcfm}  =  \sum_{i,j}\int
                           dx_1 dx_2 |M^{(\proc)}_{ij}|^2 \delta(x_1 x_2 s -
                M^2) 
 \frac{\alpha_s}{2\pi} \left[ {\cal H}^{(1)}_{\mcfm,\proc}\>f_i\!\left(x_1, \mu_F\right)
  f_j\!\left(x_2, \mu_F \right)  \notag \right. \\
  &+\left. \sum_{k}\left(
  \int_{x_1}^1\frac{dz}{z} C_{\mcfm,ik}^{(1)}(z)
  f_k\!\left(\frac{x_1}{z}, \mu_F \right)
  f_j\!\left(x_2, \mu_F \right) +\{(x_1,i)\,\leftrightarrow\,(x_2,j)\}\right)\, \right]\,.
\end{align}
After direct inspection of the \mcfm\ code, we realised that the term
$\mathcal{H}_{\mcfm,\proc}^{(1)}$ does not contain just the finite
part of the virtual corrections $H^{(1)}$, but also the terms
$-(\pi^2/12)\, \delta(1-z)$ in the coefficients $C^{(1)}_{qq}(z)$ and
$C^{(1)}_{gg}(z)$, as well as terms containing
$\ln(M^2/\mu_R^2)$. Keeping this in mind, to compute the luminosity
$\mathcal{L}_{\proc}^{(1)}$ through \mcfm, we had to perform the following
changes to the \mcfm\ code:
\begin{enumerate}
\item replace $\mathcal{H}_{\mcfm,\proc}^{(1)}$ as follows
\begin{equation}
  \label{eq:HHmcfm}
  \begin{split}
    \mathcal{H}_{\mcfm,q\bar q}^{(1)} &  \to \mathcal{H}_{\mcfm,q\bar q}^{(1)}+2C_F\left(\frac{\pi^2}{12}+\frac{3}{2}\ln\frac{Q^2}{\mu_R^2}+\frac{1}{2}\ln\frac{M^2}{\mu_R^2}-\ln^2\frac{M^2}{Q^2}\right)\,, \\
    \mathcal{H}_{\mcfm,gg}^{(1)} & \to \mathcal{H}_{\mcfm,gg}^{(1)}
    +2C_A\left(\frac{\pi^2}{12}+2\pi\beta_0\ln\frac{Q^2}{\mu_R^2}+\frac{1}{2}\ln\frac{M^2}{\mu_R^2}-\ln^2\frac{M^2}{Q^2}\right)
    \,;
  \end{split}
\end{equation}
\item modify the integrated counterterms as follows
  \begin{equation}
    \label{eq:C1mcfm-C1}
    C_{\mcfm,ij}^{(1)}(z) \to \frac{1}{1-2\alpha_s \beta_0 \tilde{L}} \, C_{ij}^{(1)}(z)\,;
  \end{equation}
\item change the factorisation scale in all PDFs from $\mu_F$ to $\mu_F \, e^{-\tilde{L}}$.
\end{enumerate}
Last, to implement the full NNLL resummation, we just rescale the weight of each event by the factor
\begin{equation}
  \label{eq:rescale-NNLL-Ltilde}
\left(1+\mathcal{F}_{\text{clust}}(R)+\mathcal{F}_{\text{correl}}(R)\right)
  \,e^{ \tilde L g_1(\as \tilde L)+g_2(\as \tilde L)+\frac{\as}{\pi} g_3(\as \tilde L)}  \,.
\end{equation}

\subsection{Matching to fixed order and theoretical uncertainties}
\label{sec:th-uncerts}

Our \mcfm\ implementation includes the matching of resummed
predictions with NLO calculations. In particular, we have implemented
the relevant contributions to the two multiplicative matching schemes
introduced in refs.~\cite{Banfi:2012jm,Banfi:2015pju}. At NLO, the
total cross section $\sigma_{\rm NLO}$ for the production of a colour
singlet, satisfying a set of kinematical cuts for its decay products,
is given by
\begin{equation}
  \label{eq:sigma-tot}
  \sigma_{\rm NLO}=\sigma^{(0)}+\sigma^{(1)}\,,
\end{equation}
with $\sigma^{(0)}$ its Born-level contribution, and $\sigma^{(1)}$ a
correction of relative order $\alpha_s$. Similarly, at NLO, the
corresponding cross section with a jet-veto $\Sigma_{\rm NLO}(\ptjv)$
is given by
\begin{equation}
  \label{eq:Sigma-NLO}
  \Sigma_{\rm NLO}(\ptjv)= \sigma^{(0)}+\Sigma^{(1)}(\ptjv)\,.
\end{equation}
For computational convenience, it is customary to introduce
\begin{equation}
  \label{eq:Sigmabar}
  \bar\Sigma^{(1)}(\ptjv)=\Sigma^{(1)}(\ptjv)-\sigma^{(1)}\,, 
\end{equation}
which implies
$\Sigma_{\rm NLO}(\ptjv)= \sigma_{\rm NLO}+\bar\Sigma^{(1)}(\ptjv)$.
We also denote by $\Sigma_{{\rm N}^{k}{\rm LL}}(\ptjv)$ the resummed
jet-veto cross section at N$^{k}$LL accuracy, again satisfying the
chosen set of kinematical cuts for the decay products of the
considered colour singlet. At this order, it has the following
expansion in powers of $\alpha_s$:
\begin{equation}
  \label{eq:resum-expanded}
  \Sigma_{{\rm N}^{k}{\rm LL}}(\ptjv)=\sigma_0+\Sigma^{(1)}_{{\rm N}^{k}{\rm LL}}(\ptjv)\,.
\end{equation}

As in refs.~\cite{Banfi:2012jm,Banfi:2015pju}, the matching is
performed at the level of the jet-veto efficiency $\epsilon(\ptjv)$,
the fraction of events that survives the jet veto. This quantity is
matched to exact NLO, as follows:
  \begin{subequations}
  \label{eq:matching}
  \begin{equation}
    \label{eq:scheme-a}
    \epsilon^{(a)}(\ptjv)=\frac{\Sigma_{\nkll}(\ptjv)}{\sigma_{\rm NLO}}
\left[1+\frac{\Sigma^{(1)}(\ptjv)-\Sigma^{(1)}_{\nkll}(\ptjv)}{\sigma_0\left(1+\delta \mathcal{L}_{\nkll}(\ptjv)\right)}\right]\,,
  \end{equation}
  \begin{equation}
    \label{eq:scheme-b}
    \epsilon^{(b)}(\ptjv)=\frac{\Sigma_{\nkll}(\ptjv)}{\sigma_0}\left[1+\frac{\bar\Sigma^{(1)}(\ptjv)-\Sigma^{(1)}_{\nkll}(\ptjv)}{\sigma_0\left(1+\delta \mathcal{L}_{\nkll}(\ptjv)\right)}\right]\,.
\end{equation}
\end{subequations}
At NLL accuracy, $\delta \mathcal{L}_{\rm NLL}=0$. At NNLL accuracy,
if we define $\left< \mathcal{L}^{(0)}\right>$ and
$\left< \mathcal{L}^{(1)}\right>$ as the integral of the luminosities
$\mathcal{L}^{(0)}$ and $\mathcal{L}^{(1)}$ in eqs.~(\ref{eq:L0})
and~(\ref{eq:L1}) respectively over the appropriate configurations of the decay
products of the colour singlet, we have
$\delta \mathcal{L}_{\nkll}(\ptjv)\equiv \left<
  \mathcal{L}^{(1)}\right>/\left< \mathcal{L}^{(0)}\right>$. Both
matched efficiencies reduce to $\Sigma_{\nkll}(\ptjv)/\sigma_{\rm NLO}$
for $\ptjv\ll M$, up to N$^{3}$LL corrections. On the other hand, for $\ptjv \sim M$, we have
\begin{equation}
  \label{eq:nlo-efficiencies}
  \epsilon^{(a)}(\ptjv)\simeq\frac{\Sigma_{\rm NLO}(\ptjv)}{\sigma_{\rm NLO}}\,,\qquad
  \epsilon^{(b)}(\ptjv)\simeq 1-\frac{\bar\Sigma^{(1)}(\ptjv)}{\sigma_0}\,.
\end{equation}
Note also that, for $\ptjv\to \infty$, both efficiencies tend to
one, as is physically sensible.

In order to estimate the theoretical uncertainties on jet-veto cross
sections, we adapt the jet-veto efficiency method of
ref.~\cite{Banfi:2015pju} to the present situation. First, our
``central'' prediction is $\epsilon^{(a)}(\ptjv)$ with
$\mu_R=\mu_F=Q=Q_0$, with $Q_0=M/2$. Then, we vary renormalisation and
factorisation scale for $\epsilon^{(a)}(\ptjv)$ in the range
\begin{equation}
  \label{eq:murf-range}
  \frac{1}{2}\le\frac{\mu_{R,F}}{Q_0} \le 2\, ,\qquad
   \frac{1}{2} \le \frac{\mu_R}{\mu_F} \le 2\,.
\end{equation}
Then, we vary the resummation scale $Q$ for $\epsilon^{(a)}(\ptjv)$ in
the range $2/3\le Q/Q_0 \le 3/2$, with $\mu_R=\mu_F=Q_0$. In practice,
we do not vary the scales continuously, but we consider only
$\mu_{R,F}=\{1/2,1,2\}Q_0$ and $Q=\{2/3,1,3/2\}Q_0$. Our uncertainty
band is the envelope of the curves obtained by fixing the considered
scales at the boundaries of the allowed range (i.e.\ 9-point scale
variation), plus $\epsilon^{(b)}(\ptjv)$ with all scales set to $Q_0$.

We then compute the total cross section $\sigma_{\rm NLO}$ by choosing
as our central prediction the one with both renormalisation and
factorisation scales set at $Q_0$. We then perform renormalisation and
factorisation scale variations in the range~\eqref{eq:murf-range}, and
constructing an uncertainty band as for the efficiency, i.e.\ using the
values of the scales at the boundaries of the allowed region (7-point
scale variation).

Last, the central value for the jet-veto cross section is defined as
the product of the central prediction for $\sigma_{\rm NLO}$ and
$\epsilon^{(a)}(\ptjv)$, and the corresponding uncertainty band is
obtained by adding the uncertainties of the total cross section and
the efficiency in quadrature.

If the total cross section is only available at leading-order, we
perform the resummation at NLL accuracy. Since we cannot normalise
resummed cross sections using $\sigma_{\rm NLO}$,
$\epsilon^{(a)}(\ptjv)=\epsilon^{(b)}(\ptjv)$. Once we have the
efficiency, we evaluate theoretical uncertainties by adding in
quadrature the uncertainties on $\sigma_{\rm LO}$ and the jet-veto
efficiency.

\section{Numerical implementation in {\bf\mcfm}}
\label{sec:mcfm-re}

In this section we give the details of the implementation of the
resummation of jet-veto effects for colour singlets in \mcfm. We
assume that the reader can successfully compile and run the \mcfm\
code, in all its operation modes. If not, the interested reader should
consult the \mcfm\ manual~\cite{mcfm}.

\subsection{Overview}
\texttt{MCFM-RE} (an acronym for Resummation Edition) is a
modification of \texttt{MCFM-8.0} to include the resummation of
jet-veto effects in colour-singlet processes up to NNLL+LL${}_R$
accuracy. The modifications are modular, as most of the resummation
effects are included through an interface to the code
\jetvheto~\cite{jetvheto}, suitably modified to become a library
linkable to \mcfm. Although a small number of modifications require us
to directly change the \mcfm\ code, these do not interfere with its
usual modes of operation. The program is available
on request. Included in the package are a README file and an
example input card.

To run \texttt{MCFM-RE}, one must simply provide a suitably modified
\mcfm\ input card. We list here the new parameters we have added or
changes made to existing parameters, described with the same
conventions and terminology as the \mcfm\ manual.
\begin{itemize}
\item {\tt file version number}. This should match the version number
that is printed when {\tt mcfm} is executed.

\begin{center}
\{blank line\} \\
{\tt [Flags to specify the mode in which \mcfm\ is run] }
\end{center}

\item \texttt{part}
  \begin{itemize}
  \item \texttt{ll}. Jet-veto resummation at LL accuracy, i.e.\ each
    event produced by \mcfm\ is reweighted with
    $\exp[\tilde L g_1(\alpha_s \tilde L)]$.
  \item \texttt{nll}. Jet-veto resummation at NLL accuracy, see
    eq.~(\ref{eq:dsigma-NLL}).
  \item \texttt{nnll}. Jet-veto resummation at NNLL accuracy, with or
    without the inclusion of small jet radius resummation (LL${}_R$),
    see eq.~(\ref{eq:dsigma-NNLL}).
  \item \texttt{lumi0}. Calculation of the luminosity $\mathcal{L}^{(0)}$ in eq.~\eqref{eq:L0}
  \item \texttt{lumi1}. Calculation of the luminosity $\mathcal{L}^{(1)}$ in eq.~\eqref{eq:L1}
  \item \texttt{nllexp1}. Expansion of the NLL resummation at order
    $\alpha_s$ (for matching).
  \item \texttt{nnllexp1}. Expansion of the NNLL resummation at order
    $\alpha_s$ (for matching).
  \end{itemize}

\begin{center}
\{blank line\} \\
{\tt [JetVHeto resummation options]}
\end{center}

\item \texttt{observable}.
  \begin{itemize}
  \item \texttt{ptj}. The default mode of the resummation, resum logarithms of
    the jet-veto.
  \item \texttt{ptj+small-r}. Available for NNLL resummations only. Include the
    effect of resumming the jet radius at leading logarithmic accuracy.
  \end{itemize}
\item \texttt{Qscale}. This parameter may be used to adjust the value
  of the \textit{resummation scale} $Q$ introduced in
  eq.~\eqref{eq:Ltilde}. It behaves in the same way as the \mcfm\ parameters
  \texttt{scale} and \texttt{facscale} do, i.e.\ if
  \texttt{dynamicscale} is \texttt{.false.}, $Q$ is set to
  \texttt{Qscale}, otheerwise $Q=\texttt{Qscale}\times \mu_0$, with
  $\mu_0$ the dynamic scale specified by the parameter
  \texttt{dynamicscale}.
\item \texttt{Rscale}. This parameter may be used to adjust the value
  of the jet-radius resummation scale.
\item \texttt{ptjveto}. The value of the jet-veto cut $\ptjv$ in units
  of {\rm GeV}.

\begin{center}
\{blank line\} \\
{\tt [Coupling rescaling in the kappa formalism]}
\end{center}

\item \texttt{kappa\_t}. The parameter $\kappa_t$ of the Lagrangian in
  eq.~(\ref{eq:BSM-model}), a.k.a the anomalous top Yukawa coupling.
\item \texttt{kappa\_b}. Anomalous bottom Yukawa coupling. 
\item \texttt{kappa\_g}. The parameter $\kappa_g$ of the Lagrangian in
  eq.~(\ref{eq:BSM-model}).
\item \texttt{interference only}. Flag to control whether to compute
  just the interference terms, e.g. the coefficient of
  $\kappa_t \kappa_g$ arising from squaring the amplitude in
  eq.~(\ref{eq:Mgg}). All other coefficients can be determined by
  setting a single $\kappa_i, i=t,g,b$ to zero.

\end{itemize}
Normally, \mcfm\ identifies whether a process is $q\bar q$- or
$gg$-initiated, and running \mcfmre\ in resummation mode does not lead
to any problems. However, in cases like process 61, in fact $WW$
production, \mcfm\ includes in the NLO correction to a
$q\bar q$-initiated process formally higher-order $gg$-initiated
contribution. As a consequence, not specifying the colour of the
initial state leads to an ambiguity that is impossible to resolve. To
avoid such problems, we have decided that, when running \mcfmre\ in
any resummation mode for ambiguous processes, the user must impose
that a process is either $q\bar q$- or $gg$-initiated, by making use
of the \mcfm\ flags \texttt{omitgg} and \texttt{ggonly}. Failure of
doing so will result in \mcfmre\ stopping and returning an error
message.

\subsection{Details of \mcfm\ implementation} 
We modify \mcfm\ version 8.0 to include the resummation of jet-veto
effects. To this end there are two pieces that we must include, the
computation of the luminosities
$\mathcal{L}_\proc^{(0)},\mathcal{L}_\proc^{(1)}$, and the Sudakov
form factor combined with the functions
$\mathcal{F}_{\rm clust},\mathcal{F}_{\rm correl}$. The computation of
the luminosities requires structural changes to \mcfm\ whereas we are
able to include the Sudakov form factor through an interface in
\texttt{src/User/usercode.f90}.

The inclusion of the Sudakov form factor is the simplest change. The
reweighting is included through the subroutine \texttt{userplotter},
\begin{lstlisting}
  interface
    function sudakov(proc, M, muR, muF, Q, as, p, jet_radius, &
         &observable, small_r, small_r_R0, ptj_veto, order)

      ....

    end function
  end interface
\end{lstlisting}
The user should not normally make changes to this function. The reweighting is
applied to all histograms, including the default \mcfm\ ones, as \texttt{wt} and
\texttt{wt2} are \texttt{intent(inout)}, so our reweighting is applied globally.
The cost of doing the reweighting here is that the cross section returned by the
main \mcfm\ program is wrong, or rather it includes only the contribution of the
luminosities and not the Sudakov exponent. To that end we include the extra
histogram \texttt{xsec}, a single-bin histogram to record the correct total
cross section for runs with the jet-veto.

To include the luminosities we have to modify the factorisation scales
of the PDFs. Instead of adding lots of switches to the default \mcfm\
integration routines, we create our own special routines
\texttt{resmNLL.f} (based on \texttt{lowint.f}) and
\texttt{resmNNLL.f} (based on \texttt{virtint.f}), which we include in
the \texttt{src/Procdep} directory along with the other default
integration routines. The changes made in \texttt{resmNLL.f} are
modest with respect to \texttt{lowint.f}, schematically
\begin{lstlisting}
  function resmNLL(r,wgt)
  use rad_tools, only: Ltilde
  implicit none
  include `types.f'
  real(dp):: resmNLLint

  ! resummation
  include `jetvheto.f'
  real(dp) :: facscaleLtilde
  real(dp) :: L_tilde_arr(1)

  L_tilde_arr = Ltilde((/ptj_veto/q_scale/), p_pow)
  L_tilde = L_tilde_arr(1)
  if (do_lumi) then
    facscaleLtilde = facscale*exp(-L_tilde)
  else
    facscaleLtilde = facscale
  end if

  call fdist(ih1,xx(1),facscaleLtilde,fx1)
  call fdist(ih2,xx(2),facscaleLtilde,fx2)

  end
\end{lstlisting}
At the beginning of each event we determine $\tilde{L}$, and the modified
\texttt{facscale} which we call \texttt{facscaleLtilde}. We then use this scale
in the computation of the PDFs. The simplicity here is that at NLL accuracy all
we need to do is change the factorisation scale and reweight, so these changes
are very modest.

To perform the same calculation at NNLL is much more involved, since
there are three separate actions that must be performed to compute the
luminosity. First, we need to cast the virtual matrix element into the
correct form for the resummation. We do this with a utility function
in the file \texttt{src/Procdep/virtfin.f}, which performs the
replacement detailed in eq.~(\ref{eq:HHmcfm}). This is carried out by
the subroutine
\begin{lstlisting}
  subroutine virtfin(p,msq,msqv)
  real(dp) :: p(mxpart, 4)
  real(dp) :: msq(-nf:nf,-nf:nf), msqv(-nf:nf,-nf:nf)
  end subroutine virtfin
\end{lstlisting}
where one must provide the array of momenta \texttt{p(mxpart,4)}, the tree level
matrix element squared \texttt{msq(-nf:nf)} and the matrix element of the
virtual corrections \texttt{msqv(-nf:nf)} (using the conventions of \mcfm).

The second contribution to the luminosities comes from the convolution of the
coefficient functions. To include this coefficient function we modify the
integrated dipole functions located inside \texttt{src/Need/dipoles.f}, adding
switches to choose between the different types of ``dipoles'' that we have added
as well as the default \mcfm\ subtraction dipole.

The third and final piece is performed in the new integration routine
\texttt{src/Procdep/resmNNLL.f}. This calls the previous two routines,
and then performs the convolutions of all coefficient functions with
the PDFs.
\begin{lstlisting}
  function resmNNLL(r,wgt)
  use rad_tools, only: Ltilde
  implicit none
  include `types.f'
  real(dp):: resmNNLLint

  ! resummation
  include `jetvheto.f'
  real(dp) :: facscaleLtilde
  real(dp) :: L_tilde_arr(1)

  L_tilde_arr = Ltilde((/ptj_veto/q_scale/), p_pow)
  L_tilde = L_tilde_arr(1)
  if (do_lumi) then
    facscaleLtilde = facscale*exp(-L_tilde)
  else
    facscaleLtilde = facscale
  end if

  !! Move contribution of collinear counterterm into the ``dipoles''
  ! AP(q,q,1)=+ason2pi*Cf*1.5_dp*epcorr
  ! AP(q,q,2)=+ason2pi*Cf*(-1._dp-z)*epcorr
  ! AP(q,q,3)=+ason2pi*Cf*2._dp/omz*epcorr
  !! all AP terms are removed, those displayed here are just schematic

  ! extract the finite part of the virtual and modify for resummation
  ! must come before subtraction to get coefficient correct in checks
  call virtfin(p, msq, msqv)

  call fdist(ih1,xx(1),facscaleLtilde,fx1)
  call fdist(ih2,xx(2),facscaleLtilde,fx2)
  call fdist(ih1,x1onz,facscaleLtilde,fx1z)
  call fdist(ih2,x2onz,facscaleLtilde,fx2z)

  end
\end{lstlisting}
In addition, we can perform the matching with fixed-order using the
same method we have used in computing the resummation. We modify the
dipoles, this time to include the terms from the expansion of the
resummation. With these one can then compute the matched distribution
up to NNLL+NLO accuracy.

\bibliographystyle{JHEP}

\bibliography{jetvetobsm.bib}

\end{document}